\documentclass{aa}

\usepackage{graphicx}

\usepackage{txfonts}

\usepackage[hidelinks]{hyperref}

\usepackage{natbib}
\bibpunct{(}{)}{;}{a}{}{,}
\usepackage{siunitx} 
\usepackage{subfigure}
\usepackage{amsmath}
\usepackage{orcidlink}

\defcitealias{bergamini2023new}{B+23a}
\defcitealias{bergamini2023state}{B+23b}
\defcitealias{bergamini2021new}{B+21}
\defcitealias{bergamini2019enhanced}{B+19}

\begin{document}

   \title{Dynamical models of cluster members to probe the total mass properties of cluster subhalos}

   \subtitle{I. A comparison with parametric strong lensing models}

   \titlerunning{Dynamical models of cluster members to probe the total mass properties of cluster subhalos}
   \authorrunning{N. Bianchetti et al.}

   \author{N. Bianchetti\orcidlink{0009-0000-5309-3341}\inst{\ref{unimi},\ref{iasf}}, C. Grillo\orcidlink{0000-0002-5926-7143}\inst{\ref{unimi},\ref{iasf}}, G. Granata\orcidlink{0000-0002-9512-3788}\inst{\ref{unimi},\ref{unife},\ref{icg}}, P. Bergamini\orcidlink{0000-0003-1383-9414}\inst{\ref{unimi},\ref{INAFBo}}, M. Meneghetti \inst{\ref{INAFBo}}, A. Mercurio\orcidlink{0000-0001-9261-7849}\inst{\ref{UniSa},\ref{INAFNa}, \ref{INFNSa}}, P. Rosati\inst{\ref{unife},\ref{INAFBo}}, E. Vanzella\inst{\ref{INAFBo}}, \and G. B. Caminha
          }

  \institute{
    Dipartimento di Fisica, Universit\`a  degli Studi di Milano, via Celoria 16, I-20133 Milano, Italy \label{unimi}    \\
    e-mail: \href{mailto:nicola.bianchetti@unimi.it}{\tt nicola.bianchetti@unimi.it}
    \and
    INAF -- IASF Milano, via Corti 12, I-20133 Milano, Italy \label{iasf}
    \and
    Dipartimento di Fisica e Scienze della Terra, Universit\`a degli Studi di Ferrara, via Saragat 1, I-44122 Ferrara, Italy \label{unife}
    \and
    Institute of Cosmology and Gravitation, University of Portsmouth, Burnaby Rd, Portsmouth PO1 3FX, UK \label{icg}
    \and
    INAF -- Osservatorio di Astrofisica e Scienza dello Spazio di Bologna, Via Piero Gobetti 93/3, I-40129 Bologna, Italy \label{INAFBo}
    \and
    Università di Salerno, Dipartimento di Fisica “E.R. Caianiello”, Via Giovanni Paolo II 132, I-84084 Fisciano (SA), Italy \label{UniSa}
    \and
    INAF -- Osservatorio Astronomico di Capodimonte, Via Moiariello 16, I-80131 Napoli, Italy \label{INAFNa}
    \and
    INFN – Gruppo Collegato di Salerno - Sezione di Napoli, Dipartimento di Fisica “E.R. Caianiello”, Università di Salerno, via Giovanni Paolo II, 132 - I-84084 Fisciano (SA), Italy \label{INFNSa}
  }
 
  \abstract{In this series of papers, we present dynamical models of cluster members in strong lensing galaxy clusters to independently probe the persistent discrepancy reported between strong lensing models and cosmological hydrodynamical simulations, in terms of total mass properties for the cluster subhalos.
  In this work, we focused our study on early-type galaxies within the clusters Abell 2744 ($z=0.309$) and MACS J0416.1$-$2403 ($z=0.397$). We took advantage of deep spectroscopic data from the Multi Unit Spectroscopic Explorer (MUSE), on the Very Large Telescope, complemented with Hubble Frontier Fields photometry. We used a pipeline based on spectral fitting to perform kinematic measurements of the line-of-sight velocity dispersion profiles of 109 cluster members, extending, in the best cases, up to approximately $15\; \mathrm{kpc}$ from the galactic centers. We modeled the cluster members assuming a dual pseudo-isothermal total mass density distribution and a Jaffe stellar mass density distribution. From the models, we inferred the values of the central stellar velocity dispersion, $\sigma_0$, and the truncation radius, $r_t$, for the galaxies in our sample. We found that $\sigma_0$ is accurately recovered for all of the cluster members and varies approximately from $40\,\mathrm{km\;s^{-1}}$ to $370\,\mathrm{km\;s^{-1}}$, while $r_t$ is reliably measured for a fraction of galaxies in our sample, with sufficiently extended radial kinematic coverage. Our dynamical models predicted line-of-sight velocity dispersion profiles that fit the measured ones better than those inferred from strong lensing models.
  To tackle the dicrepancy from the strong lensing side, we exploited the $\sigma_0$ measurements obtained from the dynamical models to calibrate the Faber-Jackson scaling relations for the cluster members in both galaxy clusters. When comparing our relations to those obtained in previous kinematics and strong lensing works, we found systematically higher normalization and compatible slope and scatter values.
  We conclude that our dynamical measurements of $\sigma_0$ and $r_t$, along with calibrated scaling relations, are more robust than previous kinematic estimates which are biased by not taking into account the effects of the PSF and of averaging inside the aperture, and should therefore be adopted as improved initial prescriptions in future strong lensing models.}

   \keywords{galaxies: cluster: individual (MACS J0416.1$-$2403, Abell 2744) - galaxies: kinematics and dynamics - galaxies: structure - galaxies: scaling relations: Faber-Jackson – dark matter - gravitational lensing: strong}

   \maketitle

\section{Introduction}
The $\Lambda$ Cold Dark Matter ($\Lambda$CDM) cosmological model describes the Universe with Cold DM particles that interact only through gravity, and a cosmological constant, $\Lambda$. From this cosmological paradigm, we know that large-scale structures in our Universe formed hierarchically, by mergers of smaller halos \citep{tormen1997rise, moore1999dark}. This process led to the formation of clusters of galaxies. Clusters are constituted of DM for $\sim 85-90\%$ of their total mass \citep{zwicky1933rotverschiebung,fabricant1986x, eyles1991distribution, smail1995gravitational, kneib1996hubble, medezinski2010detailed,grillo2015clash}. This component is further subdivided into cluster-scale DM halos and galaxy-scale DM subhalos. The remaining $10-15\%$ of the mass budget is made up of baryons.
Galaxy clusters contain thousands of galaxies embedded in the DM subhalos \citep{bertin1993structure,bacon2001sauron,gerhard2001dynamical} and are thus ideal astrophysical laboratories to study both the physical properties of DM and its interactions with baryons and galaxies, their formation and evolution.
Two different methods for studying the mass distribution of clusters and their member galaxies are cosmological hydrodynamical simulations and strong gravitational lensing (SL from here on).
Starting from some initial prescriptions about the nature of DM and its interplay with baryons, and about feedback processes, cosmological simulations follow the growth of portions of the Universe. From these simulated boxes of the Universe, one can extract information about DM halos and subhalos. Simulations show that DM halos of any mass approximately have self-similar mass density profiles, with a central cusp, usually described using the Navarro-Frenk-White \citep[NFW;][]{navarro1997universal} or Einasto \citep{einasto1965construction} distributions. Any significant discrepancy between the predictions of simulations and observations may imply that the formation of structures does not proceed as predicted by the $\Lambda$CDM model, or that some of the initial prescriptions in the simulations are inaccurate or incomplete.
Strong lensing models measure the mass properties of DM halos by reconstructing the positions of multiple images of lensed sources that lie behind an observed massive object, such as a galaxy cluster. Strong lensing is particularly effective in cluster cores, where multiple images of background sources are typically observed. However, some degeneracies critically limit the accuracy of SL in reconstructing the mass structure of galaxy cluster subhalos \citep{meneghetti2017frontier, bergamini2019enhanced}. These degeneracies stem from the fact that SL is sensitive to the total mass distribution on scales typically larger than those of a single cluster galaxy, and they can be broken by exploiting independent measurements of the stellar velocity dispersion of cluster members \citep{falco1985model, treu2004massive,meylan2006gravitational,barnabe2009two}.
In state-of-the-art parametric SL models, cluster members are typically modeled with the spherical version of the dual pseudo-isothermal elliptical (dPIE) total mass density distribution \citep{grillo2015clash}. The spherical version of a dPIE resembles a Singular Isothermal Sphere (SIS), but it presents a core radius, $r_c$, and a truncation radius, $r_t$, which define the radial scales of the transition from a cored, $\rho\propto r^{0}$, to an isothermal, $\rho\propto r^{-2}$, to a $\rho\propto r^{-4}$ profile. The dPIE mass density profile is defined as
\begin{equation}
    \label{dPIE}
    \rho(r)=\frac{\rho_0}{\left(1+\frac{r^2}{r_c^2}\right)\left(1+\frac{r^2}{r_t^2}\right)}\,,
\end{equation}
where $\rho_0=\frac{\sigma_0^2}{2 \pi G}\frac{r_c+r_t}{r_c^2 r_t}$ and $\sigma_0$ is the central velocity dispersion value. After assuming a vanishing $r_c$, the SL models are left with two free parameters for each cluster member: $\sigma_0$ and $r_t$.
To reduce the number of free parameters, the most recent parametric SL models include power-law scaling relations for the velocity dispersion and the truncation radius of the cluster members. Moreover, state-of-the-art SL models include kinematic priors on cluster galaxies to break the aforementioned degeneracies and calibrate the scaling relations with respect to the total luminosity, $L$, of each member:
\begin{equation}
    \label{eq:FJ}
    \sigma_0=\sigma_r \left(\frac{L}{L_0}\right)^{\alpha}\,,
\end{equation}
\begin{equation}
    \label{eq:beta}
    r_t=r_r \left(\frac{L}{L_0}\right)^{\beta}\,,
\end{equation}
where $\sigma_r$ and $r_r$ are optimized reference values for $\sigma_0$ and $r_t$, of the member galaxies, and $L_0$ is a luminosity reference value.
The first equation, the Faber-Jackson \citep[F-J;][]{faber1976velocity} scaling relation, is calibrated with kinematic data-points, while the slope of the second relation is connected to the slope of the first through the Fundamental Plane \citep[FP;][]{djorgovski1987fundamental}, as $\gamma=2\alpha+\beta-1$, where $\gamma$ is connected to the tilt of the FP.
Alternatively, SL models can utilize the FP, a more accurate and complex relation, to directly link the velocity dispersion of the members to their structural parameters \citep{granata2022improved}.
In this work, we intend $\sigma_0$ as defined in the dPIE total mass density distribution. This velocity dispersion value is not to be confused with the $\sigma_0$ adopted in the FP relation, which is defined as the galaxy stellar velocity dispersion through an aperture of radius $r=R_e/8$, with $R_e$ being the effective radius.

As found by \citet{meneghetti2020excess}, cosmological hydrodynamical simulations underestimate the number of galaxy-galaxy SL events. An interpretation proposed to explain the lack of these events in simulations requires that these underpredict the number of massive and compact substructures inside clusters of galaxies \citep{meneghetti2022probability, meneghetti2023persistent}.

In order to investigate the origin of this discrepancy, in this series of works, we built dynamical models of elliptical cluster members in SL galaxy clusters \citep{binney1982dynamics, binney1982m, binney2011galactic} to test SL models and cosmological simulation claims.
This is ideal since dynamics gives a completely independent probe of the total mass distribution of these objects. 

Several Hubble Space Telescope (HST) programs provide deep observations of SL events in the cores of massive clusters, such as the Cluster Lensing And Supernova survey with Hubble \citep[CLASH;][]{postman2012cluster}, the Hubble Frontier Fields \citep[HFF;][]{lotz2017frontier} campaign, and the Reionization Lensing Cluster Survey \citep[RELICS;][]{coe2019relics}. As a follow-up, spectroscopic campaigns, such as CLASH-VLT \citep{rosati2014clash}, or spectroscopic measurements from the integral field Multi Unit Spectroscopic Explorer \citep[MUSE;][]{bacon2010society} on the Very Large Telescope (VLT), have permitted secure identifications of cluster members and precise redshift measurements for hundreds of multiply lensed sources. These data enable the construction of SL models, such as those presented in \citet{bergamini2023new, bergamini2023state} (\citetalias{bergamini2023new, bergamini2023state} from here on), and the development of detailed dynamical models of cluster members. MUSE deep spectroscopic observations can be used to measure the LOS velocity dispersion profiles for the cluster members. 
The values of the kinematics of cluster members obtained with dynamical models can then be used to calibrate the F-J scaling relations and compare them with those used in SL models, consequently probing the reliability of their conclusions.
In this first paper, we focus on the description of the Jeans dynamical models for the cluster members in our dataset. We then discuss the possibility of measuring key total mass quantities that can immediately be used to better calibrate the scaling relations adopted as priors in SL models. Improving the SL models represents a first step towards the resolution of the discrepancy.

The paper is organized as follows. In Sect.~\ref{sec:data}, we provide details on our photometric and spectroscopic data. In Sect.~\ref{sec:models}, we describe the assumptions adopted to build the dynamical models for elliptical cluster members and the main parameters that can be inferred. In Sect.~\ref{sec:results}, we discuss the results of the inference procedure performed using the dynamical models. We then present a comparison between the prediction of dynamical and SL models on the LOS velocity dispersion profiles and on the F-J scaling relations.
In Sect.~\ref{sec:conclusions}, we summarize our key findings and discuss some future prospects.

In this work, we assume a flat $\Lambda$CDM cosmological model with $H_0=70\,\mathrm{km\,s^{-1}Mpc^{-1}}$ and $\Omega_m=0.30$.
\section{Data} \label{sec:data}
In this section, we summarize the photometric and spectroscopic data sets for the two clusters of galaxies studied in this work, namely Abell 2744 and MACS J0416.1$-$2403 , hereafter A2744 and M0416, at redshifts $z=0.309$ \citep{allen1998resolving, ebeling2010x} and $z=0.397$ \citep{balestra2016clash}, respectively. Abell 2744 was first discovered in \citet{couch1984distant}, with a measured virial mass $M=4.0 \times 10^{15}\,M_{\odot}$ \citep{merten2011creation}; while M0416 was first discovered during the MAssive Cluster Survey \citep[MACS;][]{ebeling2000macs}, with a measured virial mass $M=0.9\times 10^{15}\,M_{\odot}$ \citep{balestra2016clash}.
Both A2744 and M0416 were further studied during the HFF programme \citep{lotz2017frontier}. The HFF observed a restricted sample of clusters, for 140 orbits each, providing deep imaging in seven different bands (HST/ACS F435W, F606W, and F814W; and HST/WFC3 F105W, F125W, F140W, and F160W). M0416 was also included in the CLASH-VLT survey (P.I.: P. Rosati) to obtain spectroscopic data to complement HST imaging \citep{postman2012cluster}.
The magnitudes and effective radii in the F814W and F160W bands for A2744 and M0416 were measured, together with all other structural parameters, in \citet{granata2026velocity} and \citet{tortorelli2023kormendy}, respectively, using the Python package MORPHOFIT \citep{tortorelli2023morphofit}, based on SExtractor \citep{bertin1996sextractor} and Galfit \citep{peng10}.

Spectroscopic information in the core of both clusters was obtained with the MUSE instrument, covering a spectral range between $4800\,\si{\angstrom}$ and $9300\,\si{\angstrom}$ and with a spectral resolution between $R \approx 1750$ at $4650\,\si{\angstrom}$ and $R\approx 3750$ at $9300\,\si{\angstrom}$, and a spectral sampling of $1.25\,\si{\angstrom}/\text{pixel}$. The data analyzed in this paper were obtained in wide-field mode (WFM), with a field of view of $1'\times1'$, and a pixel size of $0.2'' \times 0.2''$. In detail, for M0416, the NE and SW regions of the core were observed with MUSE for $17.1 \mathrm{h}$ \citep{vanzella2021muse} and $11 \mathrm{h}$ \citep{caminha2017refined}, respectively; while, for A2744, the core was observed with five different MUSE pointings, with total exposure times between $2-5 \mathrm{h}$ \citep{mahler2018strong,richard2021atlas}. Moreover, Adaptive Optics (AO) was utilized in 14 of the 18 exposures in the NE region of M0416 \citep{vanzella2021muse}. The effects of AO on the spectral cubes were taken into account during the MUSE data reduction procedure.
The MUSE data analyzed in our work have recently been used in \citet{mozumdar2025magnus, mozumdar2025magnus2} to study the structure, kinematics, and stellar populations of some cluster members. 

\section{Methods} \label{sec:models}
This section briefly describes how we built the dynamical models of the cluster members, using their measured LOS velocity dispersion profiles and starting from some assumptions about the galaxy symmetry, total and stellar mass distributions, and the anisotropy of the stellar orbits.
The equation for the LOS velocity dispersion profiles, along with the procedure that allows us to account for aperture and point-spread-function (PSF) effects, is discussed in detail in Appendix \ref{app:model}.
We also outline here the selection criteria we followed to define the cluster member sample, as well as all the methods employed to perform the kinematic measurements on each galaxy.
Here, we assume that the stellar mass density distribution of galaxies follows the stellar luminosity density distribution, meaning that the light emitted by the stars also traces how their mass is distributed within a galaxy (the mass-follows-light assumption from here on).

\subsection{Total and stellar mass profiles} \label{ss:profiles}
As discussed in App. \ref{app:model}, in order to model the LOS velocity dispersion profile of a galaxy using Eq. \eqref{sigma_ap}, one needs to assume a total and a stellar mass distribution. Elliptical galaxies are often modeled with a dPIE total mass density profile (see Eq. \eqref{dPIE}).
This density profile can be integrated to obtain the total mass of the galaxy within a sphere of radius $r$. Assuming a vanishing central core for the analyzed galaxies, $r_c=0$, we then obtain
\begin{equation}
    \label{dPIEmassvan}
    m(r)= \frac{2\sigma_0^2 r_t}{G} \arctan\left(\frac{r}{r_t}\right)\,.
\end{equation}
The choice of a dPIE profile for the total mass density profile of cluster members is justified, for instance, by the results obtained by \citet{grillo2015clash}, where modeling cluster members with such a total mass density distribution is shown to lead to parametric SL models that reproduce well the observed positions of several multiple images.
Finally, to build dynamical models, we need to make some assumptions on the stellar mass density distribution. The Jaffe profile \citep{jaffe1983simple} is often used for this purpose. Under the mass-follows-light assumption, it is possible to substitute the total mass of the stars inside an elliptical galaxy, $M$, with their total luminosity, $L$, to obtain
\begin{equation}
    \label{JaffeL}
    \nu_J(r)=\frac{L}{4\pi r_J^3}\frac{r_J^4}{r^2(r+r_J)^2}\,,
\end{equation}
where $r_J$ is a scale radius, related to the effective radius as $r_J=R_e/0.763$.
Equation \eqref{JaffeL} allows for the determination of the surface brightness profile of an elliptical galaxy, to be substituted into Eq. \eqref{final_sigma_ap}, thereby providing a dynamical model of the galaxy.

In the paper, we primarily focus on the dynamical results obtained assuming dPIE and Jaffe profiles for the total and stellar mass density profiles, respectively (hereafter referred to as the dPIE-J model).
To properly test the model systematics and strengthen our conclusions, in App. \ref{app:comp} we also present the results obtained by exploiting a NFW total mass density distribution and a Hernquist luminosity density distribution. The additional dynamical models are referred to as NFW-Jaffe (hereafter NFW-J) and dPIE-Hernquist (hereafter dPIE-H), respectively.

\subsection{The kinematic measurements}\label{ss:kin_measurements}
In this section, we describe the pipeline we exploited to perform the kinematic measurements on the cluster members. We then discuss the measurement details and selection procedure that was followed to obtain the measured LOS velocity dispersion profiles.
\subsubsection{The pPXF algorithm}
\begin{figure}
\centering
   \includegraphics[width=\linewidth]{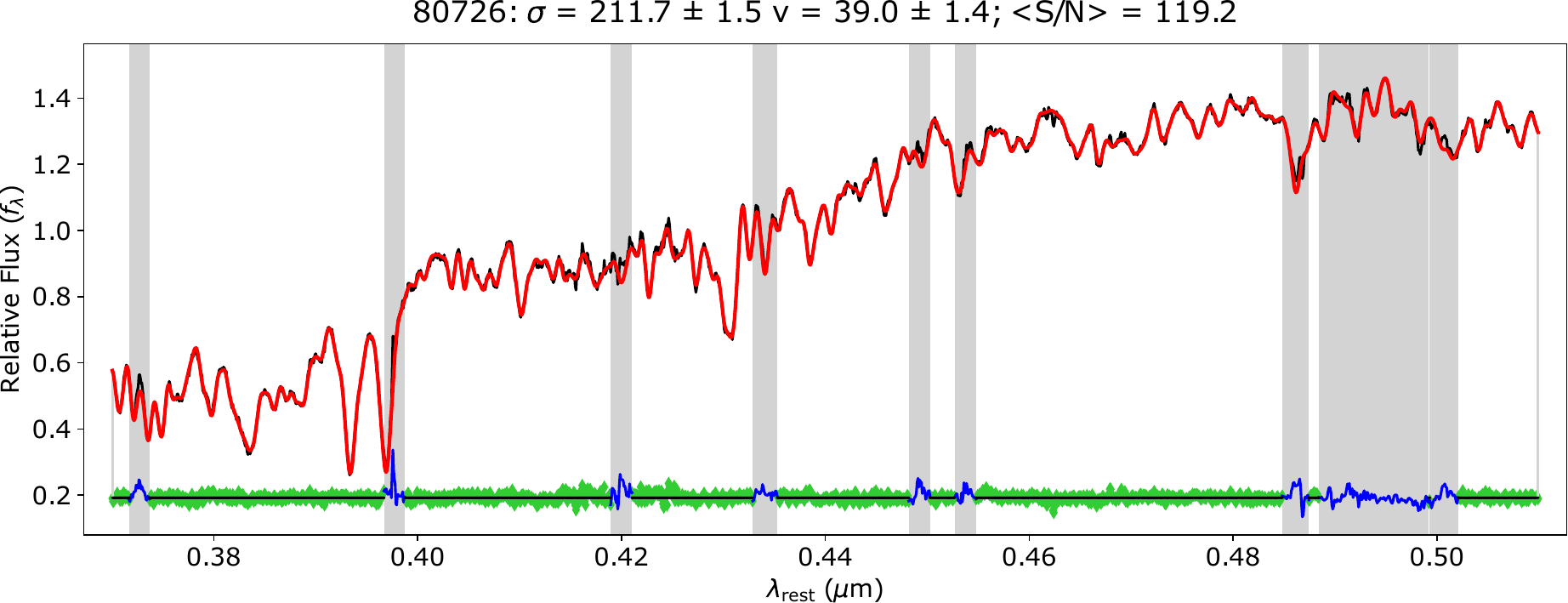}
       \caption{Spectrum of the cluster galaxy 80726 in M0416, extracted from a circular aperture with radius $r=0.5''$ centered on the galaxy luminosity center. The black line shows the measured spectrum, while the red line shows the fitted spectrum, obtained with the pPXF pipeline, from rest-frame $3700\,\si{\angstrom}$ to $5100\,\si{\angstrom}$. On the bottom, the residuals between the two spectra are shown as green dots. The gray bands highlight masked regions. On the top, the values of the extracted kinematic parameters are listed: the velocity dispersion $\sigma$ and the peculiar velocity $v$, in $\mathrm{km\,s^{-1}}$, with their corresponding uncertainties, and the average signal-to-noise ratio <S/N>.
    \label{fig:pPXF}}
\end{figure}
To perform the kinematic measurements on cluster members, we adopted the penalized pixel fitting (pPXF) software by \citet{cappellari2004parametric}. In our case, pPXF measures the first two line-of-sight-velocity-distribution (LOSVD) parameters by comparing the galaxy observed spectra with a set of 463 stellar templates we took from the second data release (DR2) of the X-shooter Spectral Library \citep{gonneau2020x}, convolved with a LOSVD. 
In addition, the algorithm utilizes additive polynomials to compensate for potential template mismatches, adjust the intensity of spectral lines, account for inaccurate spectral calibrations and dust reddening, and mitigate sky contributions to the spectrum.
The LOSVD is obtained after assuming a Gaussian parametrisation, where we only fit the first two moments of the distribution, namely the peculiar velocity and velocity dispersion values. The set of
moments of the LOSVD are obtained by the algorithm through an optimization procedure, performed
to minimize the differences between the observed and the synthetic galaxy spectra, exploiting a standard $\chi^2$ function. The first two moments of the LOSVD are thus measured, along with their corresponding uncertainties, and an average signal-to-noise value, <S/N>, is used to evaluate the reliability of the kinematic measurements. We have tested that unreliable values for the kinematic parameters are obtained when $\mathrm{S/N} < 10$. For $\mathrm{S/N} \geq 10$, the optimized values can still present a bias in the cases of spectral contamination or inaccurate fit \citep{cappellari2004parametric}.
In this work, the fits are performed on a rest-frame wavelength range of $[3700, 5100]\, \si{\angstrom}$. Further details about the kinematic pipeline we adopted are discussed in \citet{granata2026velocity}.
An example of the output of the fitting procedure is shown in Fig. \ref{fig:pPXF}. The figure presents both the observed (in black) and fitted (in red) spectra, the residuals (green dots), and the optimized LOSVD first two moments values for a cluster member in M0416. From the figure, it can be seen that the most important absorption lines used to fit the spectra in the available spectral range are the CaII K/H lines and the G band, at $3934\, \si{\angstrom}$, $3969\, \si{\angstrom}$ and $4310\, \si{\angstrom}$ (rest-frame), respectively. Galaxy emission lines, sky lines and telluric lines have been masked (grey bands in the figure), as in \cite{bergamini2019enhanced} (\citetalias{bergamini2019enhanced} from here on). As mentioned in Sect. \ref{sec:data}, the effects of AO were properly accounted for during the reduction procedure of the MUSE observations in M0416 NE. Consequently, we decided not to mask the wavelength region in which possible AO residuals lie, namely $[5800, 6000] \, \si{\angstrom}$ observed-frame. We checked that masking this interval does not influence our conclusions and that the differences that arise in the kinematic measurements with and without the mask are minor and lie below other small systematic effects that we did not consider in this work. From the figure, it is also possible to note how the depth of the MUSE observations in the cluster cores yields detailed galaxy spectra, which remarkably display a large number of characteristic absorption lines. This significantly impacts the S/N value obtained from the pPXF fitting procedure and the reliability of the measured values of the kinematic quantities and their associated errors \citep{granata2026velocity}.
\subsubsection{The galaxy sample selection procedure} \label{ss:selection}
The first step in measuring the LOS velocity dispersion profiles is the selection of a sample of elliptical galaxies within each of the two studied clusters. Starting from the full cluster member catalogs for both A2744 and M0416 \citepalias{bergamini2023new,bergamini2023state}\footnote{All of the cluster member IDs later used in the paper are obtained from the catalogs in these works.}, we require that each selected galaxy is isolated, meaning that there is no other objects closer than $1.5''$ in projection. This selection procedure is performed since some contamination from another neighboring galaxy could modify the morphology and the measured spectrum of the galaxy, substantially impacting the measured values of the kinematic parameters. Moreover, a galaxy that is surrounded by other galaxies could be influenced in its dynamics by tidal interactions. These interactions could lead to LOS velocity dispersion profiles that are much more complicated than those intended to be studied in this work. 
We then require each selected galaxy to have a $\mathrm{S/N}>30$ when a pPXF fit is performed within a circular aperture of $0.8''$, centered on its center. Since we want to infer the galaxy truncation radius, we aim to construct LOS velocity dispersion profiles with radial extensions exceeding $1''$. The requirement on the $\mathrm{S/N}$ thus favors measured profiles with at least two data points. We can still infer the $\sigma_0$ values of the few galaxies with one single measurement. Following these preliminary criteria, lists of 89 and 87 members for M0416 and A2744, respectively, were obtained for further analysis.
\subsubsection{The kinematic LOS velocity dispersion profiles}
\begin{figure}
    \centering
    \includegraphics[width=0.7\linewidth]{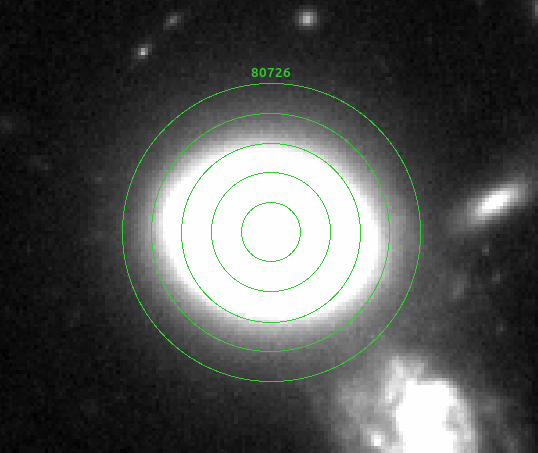}
    \caption{MUSE cube cutout of the galaxy 80726 in the cluster of galaxies M0416. The annuli of $0.5''$ width and increasing radial size, within which the LOS velocity dispersion profile was measured, are shown.}
    \label{fig:annuli}
\end{figure}
\begin{figure}
    \centering
    \includegraphics[width=0.85\linewidth]{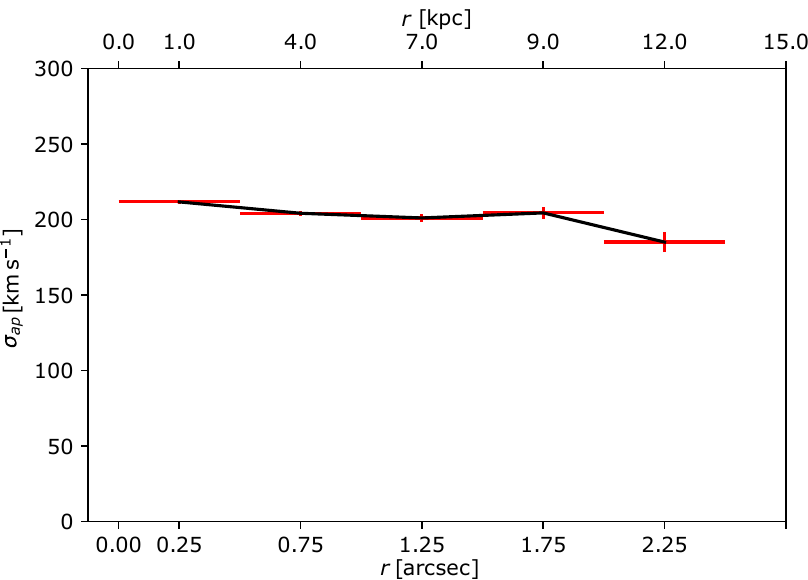}
    \caption{Measured LOS velocity dispersion profile of galaxy 80726 as a function of the radial distance from the center, in both arcsec (bottom) and kpc (top). The red lines along the horizontal axis show the uncertainty on the measurement position due to each annulus width, as shown in Fig. \ref{fig:annuli}, while the ones along the vertical axis represent the uncertainty on each measured $\sigma_{\text{LOS}}$, obtained from the pPXF pipeline.}
    \label{fig:profile}
\end{figure}

For each of the 176 preliminary selected galaxies, we then perform a pPXF fit inside annuli of $0.5''$ fixed width and of increasing radial size, to measure their LOS velocity dispersion profiles. An example of this annular subdivision is shown in Fig. \ref{fig:annuli}. The minimum width is fixed to be of the same order as the observation PSF.
A measured LOS velocity dispersion profile for a M0416 cluster member is shown in Fig. \ref{fig:profile}. In the figure, the uncertainties on $\sigma_{\text{LOS}}$ are extracted from the pPXF algorithm, while each annulus width is chosen as the uncertainty on the radius. The progressive increase in the $\sigma_{\text{LOS}}$ uncertainties with the radial distance from the galactic center is due to lower values of S/N in the outer regions of a galaxy, as a consequence of lower surface brightness.
After the measured LOS velocity dispersion profiles are built, a further selection procedure is pursued.
Exploiting HFF RGB images and MUSE spectra, it is possible to discard the profiles of those galaxies disturbed by external objects, such as foreground or background galaxies and lensed objects, or by starburst processes. Such interference can be recognized by carefully analyzing the extracted spectra for each galaxy within each annulus, searching for abrupt changes in the residuals between nearby annuli and for emission lines which are typically not expected in an elliptical galaxy spectrum.

As discussed above, pPXF is able to reliably measure the kinematic parameters and the relative uncertainties for a galaxy only if the spectrum $\mathrm{S/N}>10$. To make sure that the pPXF uncertainties are unbiased \citep{granata2026velocity} we require each of the LOS velocity dispersion measurements to have a $\mathrm{S/N}>10$. Moreover, we demand a $\mathrm{S/N}>30$ in order to proceed with the kinematic measurement in the successive annulus (e.g., in a 5-point LOS velocity dispersion profile, the first 4 points satisfy $\mathrm{S/N}\geq30$ while the last one has $10\leq\mathrm{S/N}<30$). These quite stringent limits could be relaxed in future works to allow for an extension of the sample of cluster members. Following this procedure, we obtain a total of 109 LOS velocity dispersion profiles. Within this sample, 17 profiles have three or more measured points. We highlight once again that the number of points in each LOS velocity dispersion profile is fundamental in order to reliably extract information about the truncation of the aforementioned total mass profiles. A kinematic profile that only has one or two $\sigma$ measurements is typically not extended enough radially to properly investigate its truncation. On the other hand, the central stellar velocity dispersion is reliably measured, independently of the radial extension of the LOS velocity dispersion profile, as discussed later in the paper.

Finally, from the assumptions on elliptical galaxies total and stellar mass density distributions, we made in Sect. \ref{ss:profiles}, we expect their LOS velocity dispersion profiles to decrease with $r$. Nonetheless, in our measurements we observe a fraction of increasing profiles. Such an increase could be a symptom of tidal interactions with other cluster members or a significant rotation of the galaxy. From the combination of photometric and spectroscopic analysis, it is possible to conclude that some objects classified as elliptical galaxies in the cluster member catalogs are, actually, either lenticular galaxies or fast rotators \citep{cappellari2004parametric}. A rotational contribution to the LOS velocity dispersion profile is considered as a velocity dispersion by pPXF, because kinematic measurements performed through annular apertures lead to a superposition of the red and blue shift in the galaxy spectral lines, caused by the ordered motion of the stars inside the galaxy that are respectively moving away and towards the observer. To disentangle the overlapped $\sigma-v_{\text{rot}}$ contributions, one could choose a rotation axis for these fast rotating galaxies and perform separated kinematic measurements within semi-annuli centered on that axis. To enlarge the sample of galaxies, future analyses will take this effect into account.

\subsection{Jeans dynamical models of cluster members} \label{ss:MCMC}
\begin{figure}
    \centering
    \includegraphics[width=0.8\linewidth]{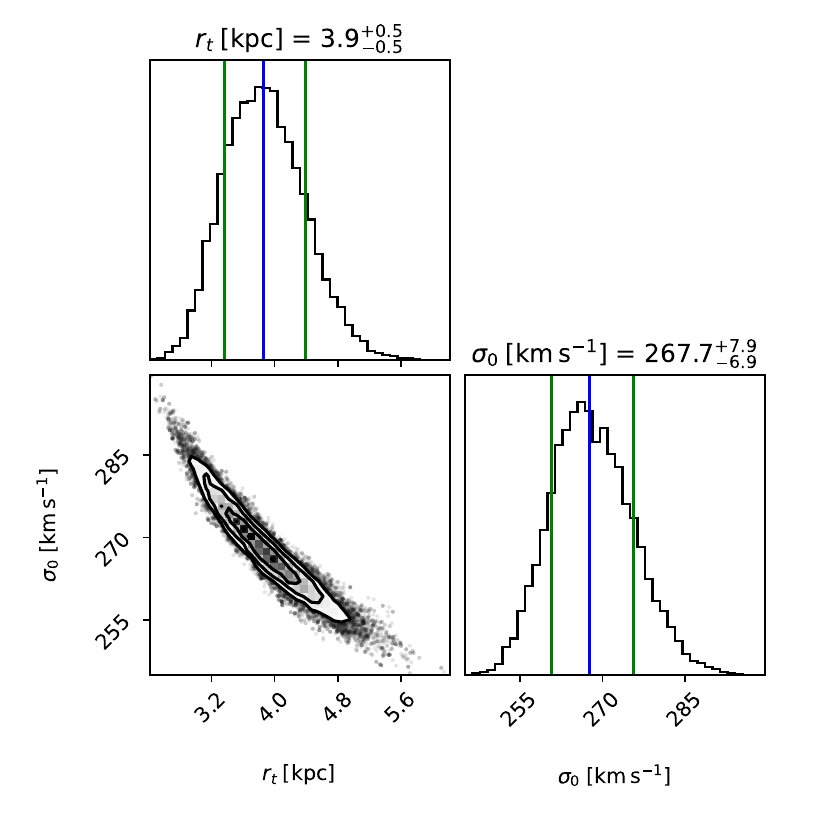}
    \caption{Corner plot obtained from the inference procedure on the cluster galaxy 83064 in M0416. The posterior probability distributions for $\sigma_0$ and $r_t$ and their correlation are shown. The median values and 16th and 84th percentiles are reported at the top and highlighted with blue and green lines.}
    \label{fig:corner}
\end{figure}

\begin{figure}
    \centering
    \includegraphics[width=0.8\linewidth]{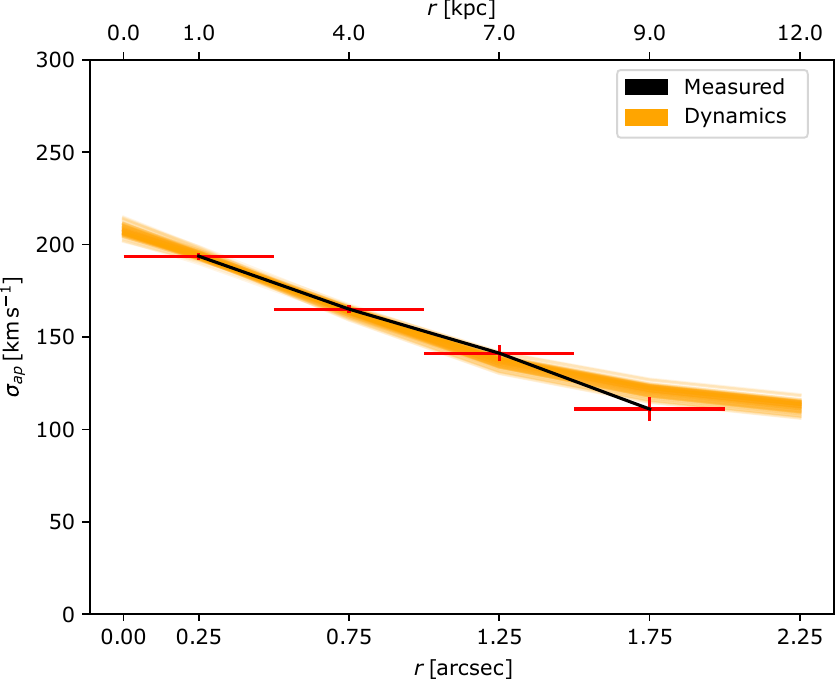}
    \caption{Comparison between measured and model-predicted velocity dispersion profiles. In black, we show the profile obtained from the kinematic measurements (red data points) of the cluster member 83064 in M0416. As orange solid lines, we plot 100 profiles obtained with the dynamical model by sampling the posterior distributions of the free parameters, drawn from the MCMC optimization and shown in Fig. \ref{fig:corner}. The distance from the galactic center is reported in arcsecs (bottom) and kpc (top).}
    \label{fig:sigmaplot}
\end{figure}

As discussed in App. \ref{app:model} and Sec. \ref{ss:profiles}, by substituting Eq. \eqref{dPIEmassvan} as the total mass density distributions and Eq. \eqref{JaffeL} as the stellar luminosity density distribution in Eq. \eqref{sigma_ap}, one obtains a Jeans dynamical model for the LOS velocity dispersion profile. Exploiting HFF photometric data (see Sect. \ref{sec:data}), one can compute the luminosity for each of the galaxies in our sample, starting from their F814W magnitude. The scale radius $r_J$ is related to $R_e$ and is thus immediately obtained.
It is then possible to combine the kinematic measurements and the dynamical model for the LOS velocity dispersion profiles to infer the values of $\sigma_0$ and $r_t$.
To perform the inference procedure, we exploited the Markov chain MonteCarlo (MCMC) optimization algorithm emcee \citep{foreman2013emcee}. We defined wide flat prior distributions for the two free parameters: $\sigma_0\in [10,1000]\,\mathrm{km\,s^{-1}}$ and $r_t\in [0.01,200] \,\mathrm{kpc}$.
For each of the sample galaxies, we randomly select a starting value for the free parameters within such priors. We then perform the MCMC optimization using chains with 4 walkers and 15000 steps each. Considering the autocorrelation time, we discard the first 1000 steps for each chain. We obtain as output the corner plots and posterior probability distributions for the free parameters. An example of the corner plots for a galaxy in M0416, parametrized with the dPIE-J dynamical model, is shown in Fig. \ref{fig:corner}.
From the obtained posterior probability distributions, one can use the dynamical models to compute the predicted LOS velocity dispersion profiles and compare them with the kinematic measurements, as shown in Fig. \ref{fig:sigmaplot}. From the figure, which compares 100 profiles obtained from the dPIE-J dynamical models (in orange) of a M0416 cluster member (ID: 83064) with the kinematic measurements (in black), it is clear that the model is good at reproducing the measured profiles.

In the following, we will focus on how the values of $\sigma_0$ inferred from the modeling can be used for calibrating a F-J scaling relation on the sample galaxies for both A2744 and M0416. We highlight that this procedure is substantially different from aperture measurements of the stellar velocity dispersion, such as the ones performed in \citetalias{bergamini2023new} and \citet{bergamini2021new} (\citetalias{bergamini2021new} from here on). The first difference is due to the fact that we measure the actual central value $\sigma_0$, whereas a measurement through an aperture yields a mean $\sigma$ value within the aperture. This effect can be seen in Fig. \ref{fig:sigmaplot}, where the $\sigma$ value measured in the first annulus (in black) is clearly lower than the central value obtained from the dynamical model (in orange). More importantly, the values measured within the apertures in \citetalias{bergamini2023new} and \citetalias{bergamini2021new} typically do not take into account the PSF of the observations, which, when convolved with the $\sigma_{\mathrm{LOS}}$, dims the central values of the profile. These effects on the measured galaxy kinematics and, subsequently, on the calibration of the F-J scaling relation are discussed in depth in the next section. 


\section{Results and discussion} \label{sec:results}
In this Section, we discuss the main results obtained through dynamical modeling and compare them with the state-of-the-art parametric SL models \citepalias{bergamini2023new,bergamini2023state} and stellar kinematic measurements \citepalias{bergamini2023new,bergamini2021new}. The comparison between the predictions of the different model parametrizations is presented in App. \ref{app:comp}.

\subsection{The different truncation radius classes}\label{ss:classes}
\begin{figure*}
    \sidecaption
    \includegraphics[width=6cm]{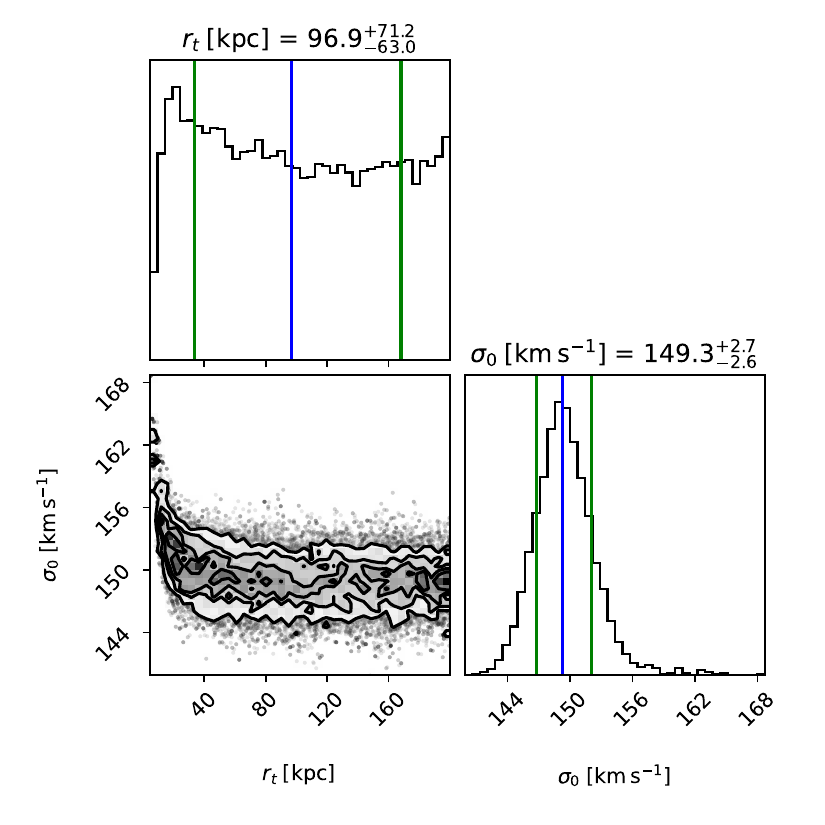}
    \includegraphics[width=6cm]{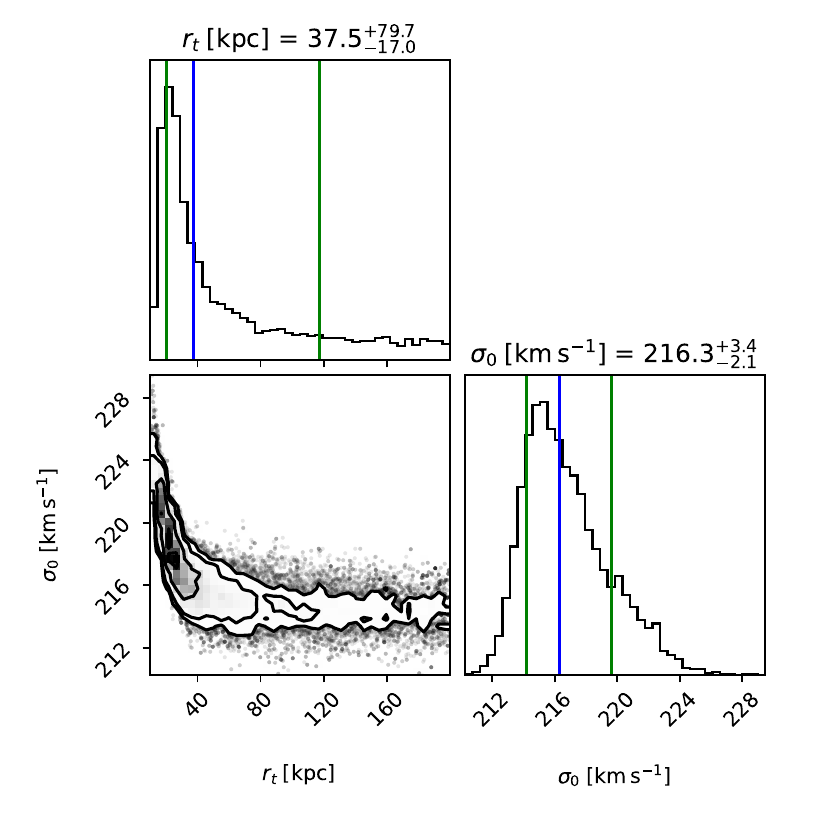}
    \caption{Corner plots of the central stellar velocity dispersion, $\sigma_0$, and truncation radius, $r_t$, for two cluster members (38729 on the left and 34423 on the right) in A2744, obtained from the inference procedure. The posterior probability distributions of $\sigma_0$ and $r_t$ and their correlations are shown. The median values and percentiles are highlighted as in Fig. \ref{fig:corner}. The $r_t$ posterior probability distributions for these galaxies are classified as `Flat' and `Tailed', respectively. The $\sigma_0$ posterior distribution is close to a Gaussian distribution for both galaxies.}
    \label{fig:flat}
\end{figure*}

Following the procedures outlined in the previous Section, we were able to obtain information about $\sigma_0$ and $r_t$ for all of the 109 selected galaxies. The measured photometric properties and the inferred values of the model parameters are listed in Tables \ref{table:res17} and \ref{table:res2p} for galaxies with, respectively, at least and less than three measured kinematic points. For all the galaxies in our sample, we obtained a posterior distribution of $\sigma_0$ similar to a Gaussian distribution. We then divide the sample of galaxies into three different classes: `Gaussian', `Tailed', and `Flat', depending on the shape of the inferred $r_t$ posterior distribution. For 97 galaxies, we obtained a flat posterior distribution for $r_t$ (see Fig. \ref{fig:flat}, left corner plot); for 5 we obtained a tailed posterior distribution (see Fig. \ref{fig:flat} right corner plot) and for 7 we obtained a simil-Gaussian distribution (see Fig. \ref{fig:corner}).
The $r_t$ median values obtained for the `Flat' galaxies depend on the selected prior, rendering the truncation estimates for `Flat' galaxies unusable.
The inability of the dPIE models to consistently obtain a reliable value for the truncation radius of a galaxy can be justified as follows:
\begin{itemize}
    \item Since all of the galaxies that have measured kinematic profiles with less than three points are classified as `Flat' (except for one, ID: 81285), we can conclude that for these galaxies the truncation radius probably lies outside of the range covered by our kinematic measurements.
    \item We noticed that the 6 galaxies, with more than two measured kinematic points, classified as `Flat', exhibit a rotational velocity contribution that increases with radial distance from their center. This leads to higher measured $\sigma_{\mathrm{ap}}$ values in the outskirts of the galaxies, subsequently causing a flattening of the measured profiles. We thus suggest that more complex dynamical models, which take into account both the stellar velocity dispersion and rotation, are needed to reliably measure the $r_t$ values for these galaxies. Nonetheless, the lack of strong rotation in the inner regions of these galaxies \citep[e.g.,][]{zhu2023manga} allows us to conclude that the $\sigma_0$ measurements of these cluster members are reliable.
    \item For the 5 galaxies classified as `Tailed', the rotation effect described above is present but does not strongly influence the measured profiles. To test this, we performed the inference procedure on some of the galaxies classified as `Gaussian', removing the last LOS velocity dispersion measurement. It appears that having a $r_t$ that is on the edge of the kinematic measurements does impact the recovered $r_t$ posterior distribution, resulting in a tail. However, this does not influence significantly the estimated $\sigma_0$ values. This suggests once again that the lack of radially extended kinematic measurements prevents us from precisely evaluating the truncation radius of some galaxies.
\end{itemize}
Thus, for all 109 sample galaxies, the $\sigma_0$ values are reliably inferred, while $r_t$ is measured for only 12 cluster members. Given the fact that we can reliably recover the values of $r_t$ only for a fraction of the sample of cluster members, no statistically significant conclusions about the compactness of these objects and its effects on the discrepancy between SL models and cosmological simulations can be made. We refer the reader to the next paper of this series in which we will increase the statistics of our analysis to directly tackle the discrepancy. On the other hand, the large sample of $\sigma_0$ values can be exploited to test the initial prescriptions of SL models, which is a first step to verify the robustness of their measurements of the compactness of DM sub-halos.
\subsection{Comparison with SL model profiles}
\begin{figure}
    \centering
    \includegraphics[width=0.9\linewidth]{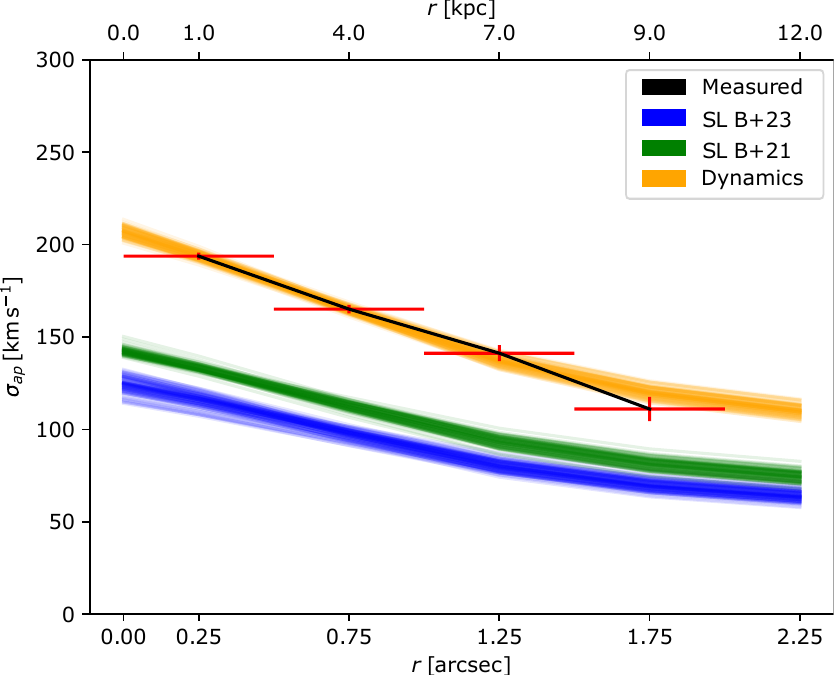}
    \caption{Comparison between the measured, SL, and dynamically modeled velocity dispersion profiles for a M0416 cluster member (ID: 83064). In black, we plot the measured kinematic profile. In orange, we show 100 profiles obtained with the dPIE-J dynamical model. In green and blue, we plot 100 profiles obtained from the calibrated and optimized F-J scaling relations from \citetalias{bergamini2021new} and \citetalias{bergamini2023state}, respectively. The uncertainties and the distance are defined as in Fig. \ref{fig:profile}.}
    \label{fig:SLcomparison}
\end{figure}

The F-J scaling relation (see Eq. \ref{eq:FJ}) adopted in galaxy cluster SL models is typically calibrated using kinematic measurements performed through apertures. This calibration is achieved by performing a MCMC optimization procedure on the normalization and slope of the scaling relation.
Once the scaling relation for a cluster of galaxies is calibrated, one can obtain information about the total mass parameters of any cluster member from its measured total luminosity.
Starting from some parametric SL models for M0416 \citepalias{bergamini2023state} and A2744 \citepalias{bergamini2023new} and their corresponding calibrated scaling relations \citepalias{bergamini2021new,bergamini2023new}, one can reconstruct the LOS velocity dispersion profiles for each of our sample galaxies and compare them with the kinematic measurements and the profiles obtained from our dynamical models.
In Fig. \ref{fig:SLcomparison}, we plot the measured LOS velocity dispersion profile (in black, with red error bars) for a galaxy (ID: 83064) in M0416 and compare it with 100 profiles obtained from the dPIE-J dynamical model (in orange) and the F-J scaling relations calibrated in the \citetalias{bergamini2021new} (in green) and \citetalias{ bergamini2023state} (in blue) SL models, sampling the posterior probability distributions of the model parameters. From the figure, it can be concluded that the dynamical model reproduces the measured profiles well, as expected, while the SL-based profiles do not provide a very good fit to the observations. In detail, both SL profiles seem to systematically predict lower values when compared to the measurements. For all galaxies in our sample, the measured profile is reproduced more accurately by the dynamical model.
The discrepancy between SL and dynamical models can be justified by the fact that the stellar kinematic measurements used to calibrate the scaling relations for SL models are performed within apertures. Such measurements typically yield lower values of $\sigma_0$, if compared to the dynamical models, because the former represent average values within apertures while the latter infer the actual central value. Most importantly, these measurements typically do not take into account the effect of the PSF which can significantly lower the measured $\sigma_{\mathrm{ap}}$ values. In their App. C, \citetalias{bergamini2019enhanced} discuss how they account for luminosity weighting and make small corrections to model aperture effects. In detail, they derive the dPIE $\sigma_0$ parameter from the aperture average line-of-sight (projected) velocity dispersion, exploiting a projection coefficient that depends on the surface brightness profile of the cluster members. This is a first correction for the luminosity weighting but does not account for the PSF. In a more recent work, \citet{granata2024cosmology} show how using weighted spectra yields an effective $\sigma_0$ value.
On the other hand, with our dynamical models, we are actually inferring the central value of the stellar velocity dispersion, $\sigma_0$, while simultaneously considering the effects of the PSF and of the annular weighting and integration procedures. We conclude that properly accounting for the PSF in our dynamical models is the primary reason why we obtain systematically higher values of $\sigma_0$. Our method therefore provides more accurate estimates of the central stellar velocity dispersion of cluster members if compared to those usually adopted in SL modeling.

\subsection{Recalibration of the Faber-Jackson scaling relations for clusters of galaxies}
\begin{figure*}
    \centering
    \includegraphics[width=0.9\linewidth]{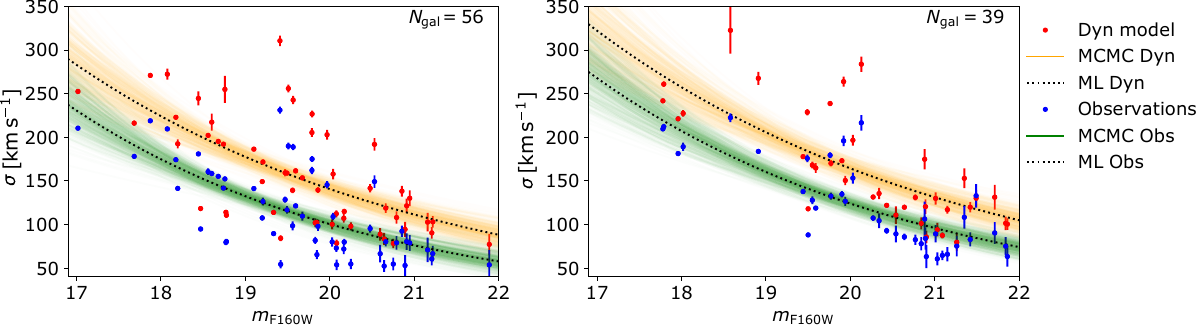}
    \caption{Comparison between kinematic measurement (`RS' sample in Tab. \ref{table:FJ}) and dynamical model F-J scaling relations for the clusters of galaxies A2744 (on the left) and M0416 (on the right). In green, we plot the MCMC best-fitting F-J scaling relations obtained from \citetalias{bergamini2023new} and \citetalias{bergamini2021new}. In orange, we present the best-fitting F-J scaling relations of our dynamical model. The measured kinematic and dynamical $\sigma$ values ($\sigma_0$ in our models) are reported as blue and red dots, respectively, with measured uncertainties as error bars. The total number of galaxies in the `RS' sample for each cluster is shown in the top-right corner. We also illustrate the maximum likelihood relations as black dotted lines.}
    \label{fig:comparisonFJ}
\end{figure*}

The results shown above suggest that observational effects may impact the calibration of the F-J scaling relations used in the SL models of galaxy clusters. Thus, the total mass estimates of the cluster sub-halos could probably be refined by combining SL models and galaxy dynamics. We thus decided to utilize the $\sigma_0$ dynamical model measurements to recalibrate the F-J relations of A2744 and M0416.
To make our scaling relations comparable to those from previous works, we reduced the sample of cluster members in \citetalias{bergamini2023new} and \citetalias{bergamini2021new} to match our sample. Our full sample of cluster members will be labeled as `FS' from here on. We also decided to remove from the kinematic sample the lenticular galaxies that we found by evaluating their morphology in the HFF images and their stellar rotation through the MUSE data. This was done because the velocity dispersion profiles we measure for these galaxies and thus the estimates of total mass parameters are probably influenced by rotation. We will refer to this restricted sample of galaxies as `RS' from here on. We exclude the BCGs, since these massive galaxies experience numerous encounters and interactions with other cluster members, which typically modify their morphology and kinematics. These effects cannot be accurately described with our dynamical model. We thus calibrate the F-J relations for 56 galaxies in A2744 and 39 galaxies in M0416 (highlighted IDs in Tables \ref{table:res17} and \ref{table:res2p}) and compare them to those obtained for the `FS' and `RS' samples, using the kinematic measurements presented in \citetalias{bergamini2023new} and \citetalias{bergamini2021new}.
To perform the calibrations, we used the $\sigma_0$ median values and 16th and 84th percentiles of the posterior probability distributions, reported in Tables \ref{table:res17} and \ref{table:res2p}, and the HFF measured $m_{\text{F160W}}$. We thus ran an MCMC optimization procedure to infer the F-J slope, $\alpha$, normalization, $\sigma_r$, and scatter, $\Delta\sigma_r$.
In Fig. \ref{fig:comparisonFJ}, we show the comparison between the F-J relations obtained with the dynamical model measurements (in orange, red dots) and the kinematic measurements (in green, blue dots) for A2744 (on the left) and M0416 (on the right), respectively, using the `RS' restricted sample of galaxies. The best fitting values of the parameters of the scaling relations obtained with our dynamical models and the ones from the aforementioned works are listed in Table \ref{table:FJ}. We also report in the first row of each sub-table the values of $\alpha$ and $\sigma_r$, obtained from the SL model optimization process. From Fig. \ref{fig:comparisonFJ}, one can see that the $\sigma$ values measured with the dynamical models properly accounting for the PSF are systematically shifted towards higher values if compared to the kinematic measurements from \citetalias{bergamini2023new} and \citetalias{bergamini2021new}. This is confirmed in Table \ref{table:FJ}, where one can see that the difference in $\sigma_r$ is of the order of $50\,\mathrm{km\,s^{-1}}$ for both clusters.
Moreover, we obtain slightly higher $\Delta \sigma_r$ values, meaning that the cluster members are not as precisely described by F-J scaling relations as found in \citetalias{bergamini2023new} and \citetalias{bergamini2021new}. This can probably be attributed to the fact that the F-J is unable to fully account for the variety of cluster members.
The $\alpha$ values obtained from SL and dynamical models are consistent. From the table, one can also see a systematic decrease in the $\alpha$ value starting from the kinematic measurements and going towards the dynamical model measurements. This is probably caused by the lack of measurements on the BCGs and on faint cluster members, which, if included, could possibly yield steeper curves.
Moreover, not including the BCGs could also be the cause of the higher SL model optimized values of $\sigma_r$, if compared to the ones obtained from the kinematic measurements.
The discrepancy between the normalization values of the F-J scaling relations could indicate that the substructures in the cores of galaxy clusters are more/less compact than previously thought. Adopting the F-J scaling relations calibrated with the dynamical results could, in fact, impact the predictions of parametric SL models on the multiple image positions and on the cluster total mass distribution. Finally, when comparing the F-J relations for the `FS' and `RS' samples, we notice that removing the lenticular galaxies slightly influences the $\alpha$ value. Indeed, these galaxies are expected to follow a F-J relation \citep{zhu2024manga, cappellari2025early} and should not impact the predicted $\alpha$ and $\sigma_r$ values or affect the reliability of SL model conclusions.

\begingroup
\renewcommand{\arraystretch}{1.15}
\begin{table}
\caption{Comparison of the parameters of the F-J scaling relation obtained from the dynamical models, the SL models optimizations, and previous literature kinematic measurements, for the clusters of galaxies A2744 and M0416.}       
\label{table:FJ}     
\centering                          
\begin{tabular}{c | c c c}
\hline
\multicolumn{4}{c} {A2744}\\
\hline
Model & $\alpha$ & $\sigma_r\,\mathrm{[km\,s^{-1}]}$ & $\Delta\sigma_r\, \mathrm{[km\,s^{-1}]}$ \\ 
\hline    
  SL: B+23a & $0.40$ & $289^{+27}_{-16}$ & $-$ \\  
  Kin `FS': B+23a& $0.33^{+0.03}_{-0.03}$ & $217^{+12}_{-12}$ & $33^{+3}_{-3}$ \\
  Kin `RS': B+23a& $0.30^{+0.04}_{-0.04}$ & $210^{+16}_{-16}$ & $36^{+4}_{-3}$ \\
  Dynamical & $0.26^{+0.04}_{-0.04}$ & $262^{+20}_{-20}$ & $47^{+5}_{-4}$ \\
  Dynamical `2p' & $0.24^{+0.06}_{-0,06}$ & $269^{+24}_{-24}$& $55^{+8}_{-6}$\\
\hline
\multicolumn{4}{c} {M0416}\\
\hline               
Model & $\alpha$ & $\sigma_r\, [\mathrm{km\,s^{-1}}]$ & $\Delta\sigma_r\, [\mathrm{km\,s^{-1}}]$\\   
\hline                       
  SL: B+23b & $0.30$ & $281^{+12}_{-19}$ & $-$\\    
  Kin `FS': B+21& $0.31^{+0.03}_{-0.03}$ & $272^{+14}_{-14}$ & $32^{+3}_{-3}$\\
  Kin `RS': B+21& $0.28^{+0.03}_{-0.03}$& $268^{+23}_{-22}$ & $30^{+4}_{-3}$ \\
  Dynamical & $0.25^{+0.04}_{-0.05}$ & $321^{+29}_{-28}$ & $41^{+6}_{-5}$ \\
  Dynamical `2p' & $0.20^{+0.05}_{-0,05}$ & $307^{+33}_{-30}$& $41^{+9}_{-6}$\\
\hline
\end{tabular}
\tablefoot{We report the slope, $\alpha$, normalization, $\sigma_r$, and scatter, $\Delta\sigma_r$, obtained from the Kinematic measurements performed in \citetalias{bergamini2023new} and \citetalias{bergamini2021new}, reported as `Kin'; Strong lensing measurements form \citetalias{bergamini2023new} and \citetalias{bergamini2023state}, listed as `SL'; and our dynamical model measurements, labeled as `Dynamical'. The `FS' and `RS' labels indicate that the full and restricted samples of galaxies from the listed papers were used to calibrate the scaling relations, respectively. The `RS' sample, further restricted to the galaxies with more than two kinematic measurements, is labeled as `2p'.}
\end{table}
\endgroup

One last potential caveat that can influence our calibration of the F-J scaling relations is the reliability of the $\sigma_0$ measurements for galaxies with only one kinematic point. A single measured point is insufficient to fit both free parameters of the dynamical models. In these cases, the models favor large $r_t\sim 200 \,\mathrm{kpc}$ values for these galaxies, substantially describing their total mass density profiles as isothermal (i.e., without a truncation). This could lead to a bias in the estimated $\sigma_0$ values of these galaxies and potentially impact the calibration of the F-J scaling relations discussed above. To test the possible impact of this effect, we performed the F-J calibration on smaller samples of galaxies with two or more measured kinematic points (`2p' samples hereafter), which consist of 20 and 35 galaxies in M0416 and A2744, respectively. The results of this procedure are reported in Table \ref{table:FJ}. From the table, it is evident that the values of $\sigma_r$ and $\alpha$ for the `2p' samples are fully consistent with those obtained for the `RS' samples. The slightly higher $\Delta \sigma_r$ value for A2744 and larger uncertainties for M0416 can be justified by the smaller number of analyzed galaxies.
As shown above, when two or more data points are available for the kinematic profile, we do not always find evidence of truncation in the total mass density profile of the galaxy. Thus, the presence of a single velocity dispersion measurement does not imply $\sigma_0$ to be necessarily underestimated.
We conclude that the impact of galaxies with a single velocity dispersion measurement on our calibrations is small, so the obtained F-J scaling relations for A2744 and M0416 can be considered robust.

\section{Conclusions} \label{sec:conclusions}
In this paper, we presented a procedure for building Jeans dynamical models of a large sample of elliptical cluster members. This is done with the aim of independently tackling the discrepancy that arises between SL models and cosmological simulations in the measured and predicted compactness of the cluster sub halos. Our models accurately fit the kinematic measurements of the LOS velocity dispersion profiles of cluster galaxies, provided certain assumptions are made about their total and stellar mass density distributions. The main dynamical models adopted a dPIE-J parametrization, and thus depended on the values of the central stellar velocity dispersion, $\sigma_0$, and truncation radius, $r_t$. We performed kinematic measurements using MUSE IF spectroscopy and HFF photometry, obtaining the LOS velocity dispersion profiles for 109 cluster members in the galaxy clusters A2744 and M0416. 
We then optimized the free parameters of the dynamical models by fitting them to the observed stellar kinematics. The $\sigma_0$ and $r_t$ posterior probability distributions were used to predict the LOS velocity dispersion profiles of each galaxy. These profiles were then compared to the kinematic measurements and to the predictions of the SL models described in \citetalias{bergamini2023new} and \citetalias{bergamini2023state}.
Moreover, we used the values of $\sigma_0$ obtained from our dynamical models to calibrate the F-J scaling relations for both A2744 and M0416, and compared them to those obtained in \citetalias{bergamini2023new} and \citetalias {bergamini2021new}.
The main results and conclusions of our analysis are summarized as follows:
\begin{enumerate}
    \item From the dynamical models, we were able to reliably infer the values of $\sigma_0$ for all the 109 galaxies included in our sample. On the other hand, it was not possible to precisely measure $r_t$ for most of them. This is due to the fact that for 92 of the 109 galaxies, the measured LOS velocity dispersion profiles do not cover a radial range extended enough to probe the truncation radius. Given this caveat, we conclude that a larger sample of cluster members and deeper kinematic measurements are needed to make statistically significant conclusions about the `too-many-lenses' problem presented in \citet{meneghetti2020excess}.
    \item The measured LOS velocity dispersion profiles are fitted more accurately by the dynamical models than by the state-of-the-art SL models presented in \citetalias{bergamini2023new} and \citetalias{bergamini2023state}.
    \item When comparing our best-fitting F-J scaling relations to those calibrated in \citetalias{bergamini2023new} and \citetalias{bergamini2021new} for the same two clusters of galaxies, we found that the measured values for the slopes, $\alpha$, and scatter, $\Delta \sigma_r$, are compatible, but we obtained higher normalization values, $\sigma_r$. This is likely due to the incorporation of the PSF and aperture averaging effects in our dynamical models, which were not accounted for in \citetalias{bergamini2023new} and \citetalias{bergamini2021new}.
\end{enumerate}
We later extended our analysis in the Appendices. In App. \ref{app:comp} we tested the systematics of the models and verified the robustness of our claims by building two additional dynamical model parametrizations, adopting the NFW total mass density distribution and the Hernquist stellar mass density distribution. We found that parameterizations do not strongly influence the results of our analysis. Interestingly, we notice a preference towards the dPIE total mass density distribution. Since cosmological simulations typically model the cluster members adopting the NFW total mass profile, our results seem to suggest that the simulation initial prescriptions could be refined as well, possibly alleviating the discrepancy.
Moreover, in App. \ref{app:A} we found that assuming realistic values for the stellar orbit anisotrpy parameter in our dynamical models does not have a significative impact on our conclusions.

Being able to measure $\sigma_0$ and $r_t$ for a larger number of galaxies could allow us to independently probe the compactness of their DM halos and reach statistically significant conclusions on the discrepancy between SL models and cosmological hydrodynamical simulations. We suggest that adopting dynamical model measurements to recalibrate the scaling relations of cluster members in SL models could improve the accuracy of their description of the mass properties of the galaxies, possibly alleviating the discrepancy (Bianchetti et al., in prep.).
Moreover, independent measurements of the compactness of cluster members could serve as initial prescriptions for cosmological hydrodynamical simulations. This could lead to a better understanding of the DM nature and its interactions with baryons, as well as galaxy formation and evolution within dense environments.

In the next paper of this series, the methods presented in this work will be applied to an extended sample of galaxy clusters to draw some quantitative conclusions about the `too-many-lenses' problem. We will also show that deeper MUSE observations in the cores of A2744 and M0416 would allow us to extend the measured LOS velocity dispersion profiles and reliably probe $r_t$ in a larger sample of cluster members, enabling us to further test the discrepancy. Moreover, we suggest that more complex dynamical models that consider both the ordered and disordered motions of stars inside the galaxies could allow us to measure $r_t$ more accurately. For instance, future analyses could adopt some Jeans Anisotropic Modeling techniques \citep[JAM;][]{cappellari2008measuring, cappellari2020efficient}.

\begin{acknowledgements}
     We acknowledge financial support through grant PRIN-MIUR 2020SKSTHZ.
\end{acknowledgements}

\bibliographystyle{aa}
\bibliography{biblio.bib}
\begin{appendix}

\section{The LOS velocity dispersion profiles}\label{app:model}
In this Appendix, we present the mathematical derivation of the equation of the LOS velocity dispersion profiles used in this work.

The collisionless Boltzmann equation (CBE) is commonly adopted when studying galaxy dynamics \citep{binney1982m, binney2011galactic}. This equation describes the evolution of the probability density distribution $f(\boldsymbol{x}, \boldsymbol{v}, t)$ of a collisionless system of many particles, dynamically regulated by a gravitational potential $\Phi(\boldsymbol{x}, t)$. As shown in \citet{mo2010galaxy}, \citet{ binney2011galactic}, and \citet{bertin2014dynamics}, it is possible to simplify the full CBE to obtain the Jeans equation. It must be said that solving the Jeans equation does not guarantee that that there is a non-negative distribution function associated to the Jeans solution. On the other hand, the Jeans modeling makes the analysis much more straight forward than working with the full CBE and is thus often adopted for galaxy dynamical modeling. Elliptical galaxies are typically approximated as spherical bodies \citep{binney1982m, agnello2014dynamical}, to simplify the models. Under this approximation, it is possible to link the Jeans equation to observable quantities. After projecting the quantities on the 2D sky through the Abel's transform and weighting with the projected light profile, one can thus obtain the LOS velocity dispersion of stars inside an elliptical galaxy, as
\begin{equation}
\label{sigma_LOS}
\sigma_{\mathrm{LOS}}^2(R)=\frac{2 G}{I(R)}\int_R^{\infty} \frac{m(r)\nu(r)}{r^2}\left(\sqrt{r^2-R^2}+ k_{\beta}(r,R)\right)\, \mathrm{d}r\,,
\end{equation}
where $G$ is the gravitational constant, $I(R)$ is the galaxy surface brightness profile, $\nu(r)$ is the stellar luminosity density used to describe the stellar mass density profile, $m(r)$ is the total mass value of the galaxy at a given distance, $r$ and $k_{\beta}(r,R)=\int_R^r \frac{(2 r'^2-3R^2)\beta(r') J_{\beta}(R,r')}{r' \sqrt{r'^2-R^2}}\mathrm{d}r'$, in which $J_{\beta}(r,r')=\exp\left[\int_{r}^{r'} \frac{2 \beta(s)}{s}\mathrm{d}s\right]$ and $\beta(r)=1- \frac{\langle v_{\theta}^2\rangle}{\langle v_r^2\rangle}$ is the anisotropy parameter that quantifies the level of anisotropy of the stellar orbits inside a galaxy. 
Eq. \eqref{sigma_LOS} thus implies that it is possible to link the kinematic measurements of a galaxy with its total and stellar mass distributions. The mass-follows-light assumption implies that we assume a constant $M_{*}/L$ ratio, which makes it easier to link the luminosity and stellar mass density distributions.

As described in the main body of the paper, we perform kinematic measurements on the cluster members through an IF spectrograph within annular apertures. As shown in \citet{granata2026velocity}, one can obtain the LOS velocity dispersion within an aperture, $\sigma_{\text{ap}}$, by weighing the LOS velocity dispersion in Eq. \eqref{sigma_LOS} with the stellar light distribution within the aforementioned aperture, as
\begin{equation}
    \label{sigma_ap}
\begin{split}
\sigma_{\mathrm{ap}}^2(R_1,R_2)=\frac{1}{L(R_1,R_2)}\int_{R_1}^{R_2} 2 \pi R\, I(R)\,\sigma_{\mathrm{LOS}}^2(R)\, \mathrm{d}R=\\
    =\frac{4 \pi G}{L(R_1,R_2)}\int_{R_1}^{R_2} R \int_{R}^{\infty} \frac{\nu(r) m(r)}{r^2}\left[\sqrt{r^2-R^2}+ k_{\beta}(r,R)\right]\, \mathrm{d}r\,\mathrm{d}R\,,
\end{split}
\end{equation}
where $R_1$ and $R_2$ are, respectively, the internal and external radii of the annulus within which the measurement is performed, and $L(R_1,R_2)=\int_{R_1}^{R_2} 2\pi R' I(R')\, \mathrm{d}R'$ is the galaxy luminosity within the same annulus.
In order to simplify Eq. \eqref{sigma_ap}, we assume isotropic stellar orbits for the analyzed galaxies, i.e. $\beta(r)=0$. This assumption is justified by the fact that violent collisionless relaxation \citep{lynden1967statistical} is expected to lead to isotropic orbits in the galactic cores (that is, the main region on which we focus our kinematic measurements) and radially-biased orbits in the outermost regions \citep{osipkov1979spherical, merritt1985spherical, cappellari2025early}. Under this assumption, we obtain 
\begin{equation}
    \label{final_sigma_ap}
    \sigma_{\mathrm{ap}}^2(R_1,R_2)=\frac{4 \pi G}{L(R_1,R_2)}\int_{R_1}^{R_2} R \int_{R}^{\infty} \frac{\nu(r) m(r)}{r^2}\sqrt{r^2-R^2}\,\mathrm{d}r\, \mathrm{d}R\,.
\end{equation}
In order to make these model-predicted LOS velocity dispersion profiles comparable with the observed ones, we need to include the effect of the PSF.
To take the PSF into consideration, we sample the LOS velocity dispersion in Eq. \eqref{sigma_LOS} and the surface brightness of the galaxy, $I(R)$, over a plane, $\sigma_{\text{LOS}}(x,y)$ and $I(x,y)$. The extension of the surface on which we perform the sampling is significantly larger than the galaxy angular dimension on the sky, ensuring that the PSF is properly taken into account and that we are not prematurely truncating the models. The 2D profiles are obtained by interpolating over the 1D profiles and exploiting the circular symmetry assumption. The sampling is performed over a grid of $0.01''\times 0.01''$ pixels. The 2D profiles are then convolved with the PSF, modeled as a 2D Gaussian kernel with Full-Width-at-Half-Maximum (FWHM) of $0.6''$ and $0.8''$ for M0416 and A2744, respectively. Since the PSF acts on the surface brightness profile of each galaxy, the convolution is performed before the light weighting procedure that leads to Eq. \eqref{final_sigma_ap}.
In order to obtain $\sigma_{\text{ap}}(R_1,R_2)$, the light weighing is performed by dividing the sum of the contribution of each pixel lying within the annulus $[R_1,R_2]$ of the 2D map $ I(x, y)\times\sigma^2_{\text{LOS}}(x, y) $ by the sum of the contribution of each pixel within the same annulus of the 2D map $I(x, y)$. The whole procedure can be summarized as
\begin{equation}
    \label{sigma_ap_PSF}
     \sigma_{\mathrm{ap, PSF}}^2(R_1,R_2)=\frac{\sigma^2_{\mathrm{PSF}}(R_1,R_2)}{L_{\mathrm{PSF}}(R_1,R_2)},
\end{equation}
with
\begin{equation}
    \label{sigma_PSF}
    \sigma^2_{\mathrm{PSF}}(R_1,R_2)= \sum_{R=R_1}^{R_2} 2 \pi R\, (I(x,y)\ast \mathrm{PSF})\,(\sigma_{\mathrm{LOS}}^2(x,y)\ast \mathrm{PSF})\,,
\end{equation}
and
\begin{equation}
    \label{L_PSF}
    L_{\mathrm{PSF}}(R_1,R_2)=\sum_{R=R_1}^{R_2} 2\pi R (I(x,y)\ast \mathrm{PSF})\,,
\end{equation}
and where $R=x^2+y^2$ is the distance of each pixel from the galaxy center.
This procedure follows Eq. \eqref{final_sigma_ap} and takes into account the fact that we work on a 2D pixel map, practically substituting integrals with sums.
From the iteration of Eq. \eqref{sigma_ap_PSF} on annular apertures of increasing radial size, one can thus obtain a LOS velocity dispersion profile for each galaxy.
It is essential to note that when building our dynamical models, we assume that the stars in the cluster members exhibit no ordered motions. This is a good assumption in the central regions of almost all early-type galaxies (ETGs). A rotational velocity contribution would require different assumptions and complicate the modeling. We discussed in the main body the possible effects of a rotational velocity contribution to the measured LOS velocity dispersion profiles and the impact of rotation on the recovered total mass parameter values for the cluster members.

\section{Testing the dynamical model systematics}\label{app:comp}
In this Appendix, we test the robustness of our dynamical modeling and of our conclusions by adopting different parametrizations for the total and stellar mass density distributions of the cluster members.

\begin{figure}
    \centering
    \includegraphics[width=0.85\linewidth]{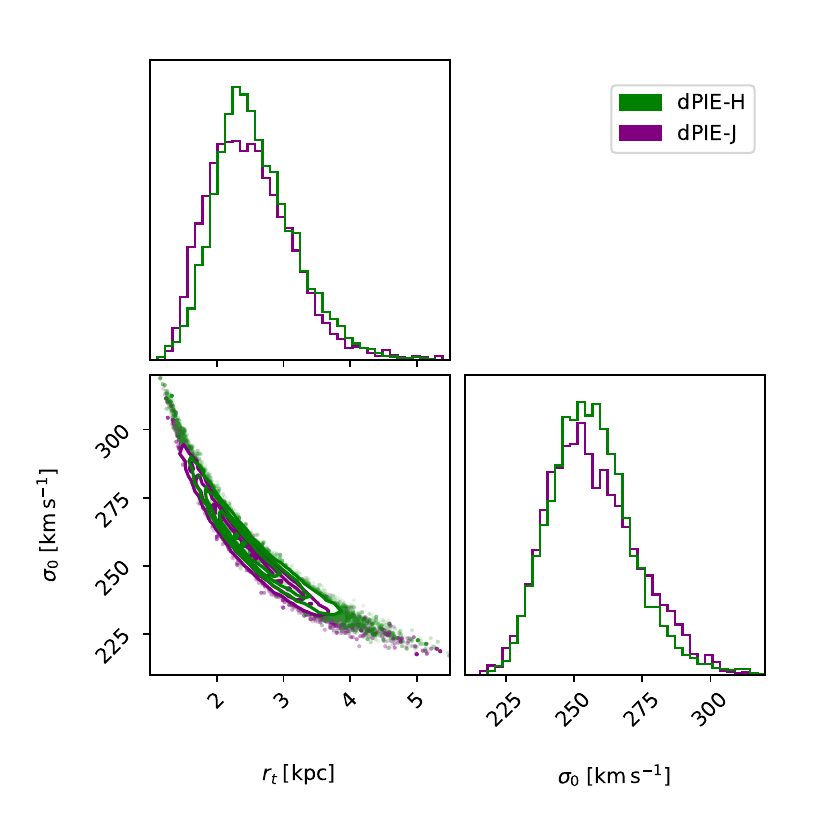}
    \caption{Comparison of the posterior probability distributions of $\sigma_0$ and $r_t$ obtained from the inference procedure performed on different dynamical model parameterizations for a A2744 cluster member (ID: 38010). We show the dPIE-J and dPIE-H posterior distributions in purple and green, respectively.}
    \label{fig:comparisonJH}
\end{figure}

\begin{figure}
    \centering
    \includegraphics[width=0.8\linewidth]{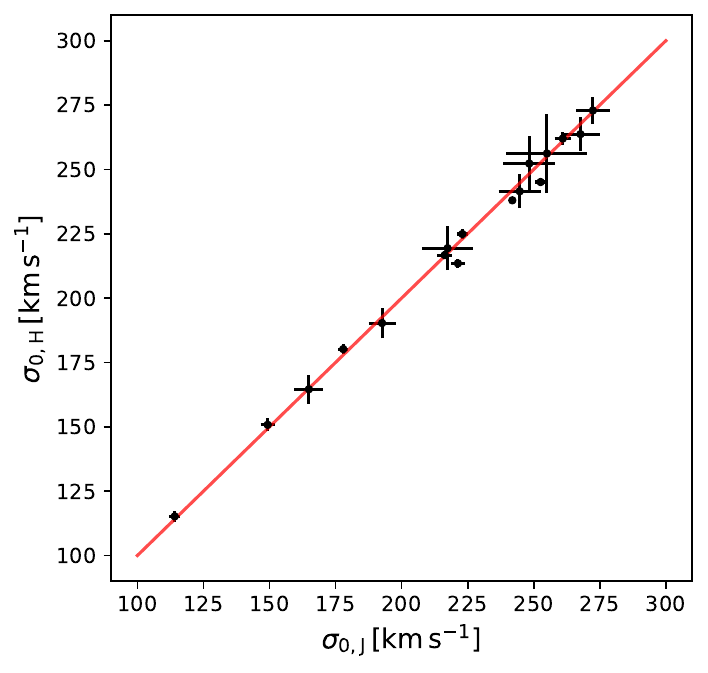}
    \includegraphics[width=0.8\linewidth]{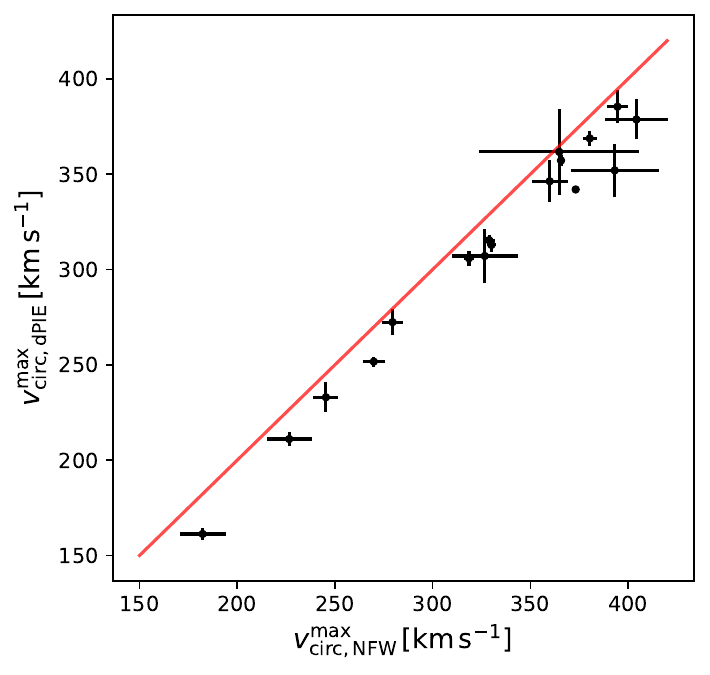}
    \caption{Comparison of the values of the central stellar velocity dispersion and maximum circular velocity, obtained through different dynamical model parameterizations. Top: correlation between the central stellar velocity dispersion values $\sigma_0$, obtained from the dPIE-J and dPIE-H dynamical models. Bottom: correlation between the maximum circular velocity $v_{\mathrm{circ}}^{\mathrm{max}}$, estimated from the NFW-J and dPIE-J dynamical models. The values and errors are obtained as the median and 16th and 84th percentiles of the posterior probability distributions, inferred through an MCMC analysis. The red line in each plot shows the one-to-one relation.
}
    \label{fig:comparisonmodels}
\end{figure}

All the results discussed in the main body of the paper are obtained by assuming a dPIE-J parametrization to dynamically model the cluster members. Cosmological hydrodynamical simulations often represent the DM structures and substructures in terms of a NFW mass density profile \citep{navarro1997universal}. The total mass of a NFW profile within a sphere of radius $r$ can be obtained, integrating the mass density profile, as 
\begin{equation}
    \label{eq:NFWmass}
    m(r)=4\pi\rho_0 r_s^3 \left[\text{ln}\left(\frac{r_s+r}{r_s}\right)-\frac{r}{r_s+r}\right],
\end{equation}
where $\rho_0$ is the central density and $r_s$ is a scale radius proportional to the virial radius of the halo, $r_{\text{vir}}=c\times r_s$, through the concentration parameter $c$.
Dynamical models of cluster members built from Eqs. \eqref{dPIEmassvan} or \eqref{eq:NFWmass} depend on different free parameters. This implies that the measured values of these parameters obtained by adopting either a dPIE or a NFW profile cannot be compared immediately. 
However, it is always possible to analyze the maximum circular velocity values obtained with the two mass distributions. For a spherical dPIE profile with a zero core radius, the maximum circular velocity of a galaxy is related to its velocity dispersion as
\begin{equation}
    \label{sigma_vcirc}     
    v_{\text{max,dPIE}}^{\text{circ}}=\sqrt{2}\,\sigma_0\,.
\end{equation}
Knowing that the maximum circular velocity for a NFW distribution is obtained when $r=\alpha r_s$, with $\alpha=2.16$, one obtains
\begin{equation}
    \label{NFW_sigma_0}
     v_{\text{max, NFW}}^{\text{circ}}=1.64\, r_s\sqrt{G \rho_0 }\,.
\end{equation}
An alternative choice for the stellar mass density profile is the Hernquist profile \citep{hernquist1990analytical}, which, under the mass-follows-light assumption, is
\begin{equation}
    \label{HernquistL}
    \nu_H(r)=\frac{L}{2\pi r_H^3}\frac{r_H^4}{r(r+r_H)^3}\,,
\end{equation}
where $r_H$ is a scale radius, related to the effective radius as $r_H=0.551\, R_e$.
We thus built additional dynamical models with the dPIE-H and NFW-J parameterizations and used them to test the systematics of our dynamical modeling and verify the robustness of the measurements of $\sigma_0$ and $r_t$, as well as our claims. We highlight that using the NFW profile is a first test of the compatibity of the initial prescriptions adopted in simulations with the actual kinematic measurements.
We thus ran the inference procedure (see Sect. \ref{ss:MCMC}) for the 17 galaxies with more than two measured points in their LOS velocity dispersion profiles using the new models. We used the same priors as in the main text for the dPIE-H model and imposed flat priors: $\rho_0\in [10^8,10^{12}]\,\mathrm{M_{\odot}\,kpc^{-3}}$ and $r_s\in [0.01,100] \,\mathrm{kpc}$ for the NFW-J parametrization.
We then compared the estimated values of the parameters with those previously obtained with the dPIE-J model.
The comparison between the dPIE-J and dPIE-H parameterizations is straightforward, because they only differ in the assumed $\nu(r)$, which is fully described by the photometric data and does not need new free parameters. So, we directly compare the measured $\sigma_0$ and $r_t$ posterior distributions. An example of this comparison for one of the sample galaxies is shown in Fig. \ref{fig:comparisonJH}. 
From the figure, we notice that the posterior distributions are almost identical. In detail, for 14 of the 17 sample galaxies, the inferred $\sigma_0$ and $r_t$ posteriors completely overlap and are nearly indistinguishable. For the three remaining galaxies, the posterior distributions are still consistent but exhibit small differences, with the dPIE-J model favoring a slightly higher $\sigma_0$ and a slightly lower $r_t$ value. In the top panel of Fig. \ref{fig:comparisonmodels}, we present the measured $\sigma_0$ values with the dPIE-J and dPIE-H dynamical models for all 17 of the aforementioned galaxies. The red line in the figure represents the one-to-one relation. From the figure, it is clear that the measurements obtained with the two different parameterizations are consistent, given the uncertainties, and are located very close to the one-to-one relation.
Moreover, if one estimates the total subhalo mass ($M_{\mathrm{sub}}$) of the galaxies, the values obtained with the two different dynamical models are fully consistent. This implies that the two free parameters, $\sigma_0$ and $r_t$, can have slightly different values, but some degeneracy ultimately leads to the same measured $M_{\mathrm{sub}}$.
In summary, a change in the stellar mass density distribution does not seem to affect the capability of the dynamical models to measure the values of $\sigma_0$ and $r_t$ and reconstruct the observed velocity dispersion profiles.
Additionally, a $\chi^2$ test can be performed to assess the goodness of fit of the two models in reproducing the measured profiles. This is achieved by generating LOS velocity dispersion profiles for all dynamical models, utilizing the 14000 accepted steps in each MCMC chain, and by searching for the lowest $\chi^2$ value to identify the model that best reproduces the measurements. The results of this analysis are shown in Table \ref{table:chisqu}. Out of the 12 galaxies for which the $\chi^2$ analysis prefers a dPIE model, there is no clear preference within the two $\nu(r)$ parameterizations. In detail, 5 galaxies are better described by a Jaffe profile, while the remaining 7 are better fitted by a Hernquist profile, but the $\chi^2$ values are always similar.

\begin{figure}
    \centering
    \includegraphics[width=0.9\linewidth]{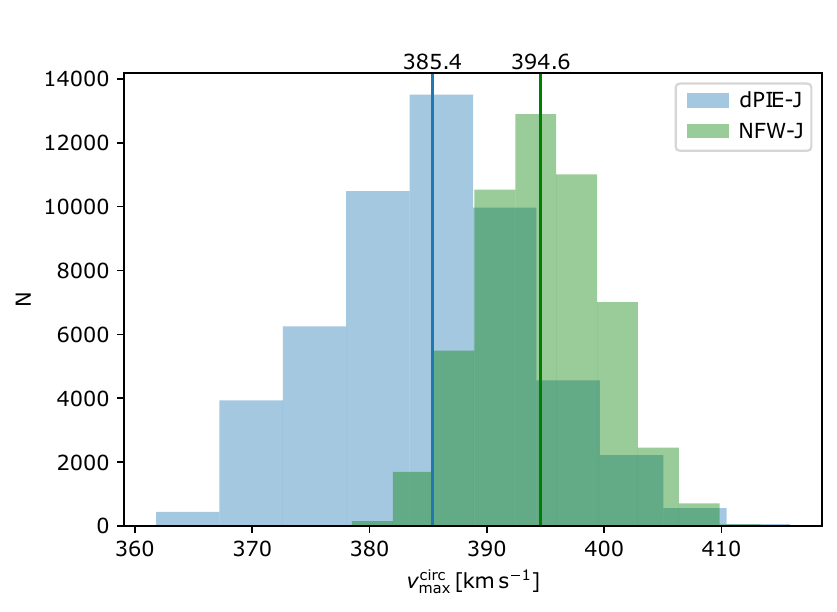}
    \caption{Comparison of the $v_{\text{max}}^{\text{circ}}$ probability distributions obtained from different dynamical model parameterizations for a cluster member (ID: 36527) in A2744. The distributions and their median values, obtained from the dPIE-J and NFW-J models, are shown in cyan and green, respectively.}
    \label{fig:comparisonNFW}
\end{figure}

The comparison between the dPIE-J model and the NFW-J model is less obvious, since the inference procedure is run on different free parameters.
To perform a meaningful comparison, we considered the value of the maximum circular velocity of the galaxies. An example of the comparison between the $v_{\text{max}}^{\text{circ}}$ distributions for one of the galaxies in our sample is shown in Fig. \ref{fig:comparisonNFW}. From the figure, it is clear that the distributions of the $v_{\text{max}}^{\text{circ}}$ values obtained with the two parameterizations are of the same order of magnitude. We evaluated the level of agreement of the maximum circular velocity distributions through the $z_{\mathrm{score}}$.
The results are reported in Table \ref{table:res17}. The distributions of $v_{\text{max}}^{\text{circ}}$ are consistent (i.e., $z_{\mathrm{score}}\leq2.33$) for 11 of the 17 sample galaxies with more than two measured kinematic $\sigma$ values. The inconsistency for the remaining 6 galaxies is related to the small uncertainties of the recovered $v_{\text{max}}^{\text{circ}}$ distributions. One can also see in Table \ref{table:res17} that the galaxies with the largest $z_{\mathrm{score}}$ values are those for which we measure large values of $r_t$. On the bottom of Fig. \ref{fig:comparisonmodels}, we show the correlation between the measurements of $v_{\text{max}}^{\text{circ}}$ obtained with the NFW-J and dPIE-J models. The red line represents the one-to-one relation. From the figure, it is evident that the NFW-J model typically predicts slightly higher $v_{\text{max}}^{\text{circ}}$ values compared to the dPIE-J model, probably because of the different slopes of the total mass density distributions. One can also observe that the measured values exhibit a small scatter around the one-to-one relation, indicating that the two dynamical models yield similar predictions for $v_{\text{max}}^{\text{circ}}$.
The consistency between the two parameterizations implies that our measurements are robust and do not depend significantly on the assumed total mass distribution. It is not possible to directly compare the values of the truncation of the galaxies, since there is no straightforward relation between the two scale radii, $r_t$ and $r_s$. Interestingly, the NFW-J model is always able to infer a finite value for both $\rho_0$ and $r_s$. This is probably due to the fact that the scale radius falls at smaller distances from the galactic center than $r_t$ and thus can be constrained with less extended kinematic measurements.
From Table \ref{table:chisqu}, it is also possible to see that the $\chi^2$ analysis prefers a dPIE over an NFW model for 12 of the 17 galaxies.

\begin{figure}
    \centering
    \includegraphics[width=\linewidth]{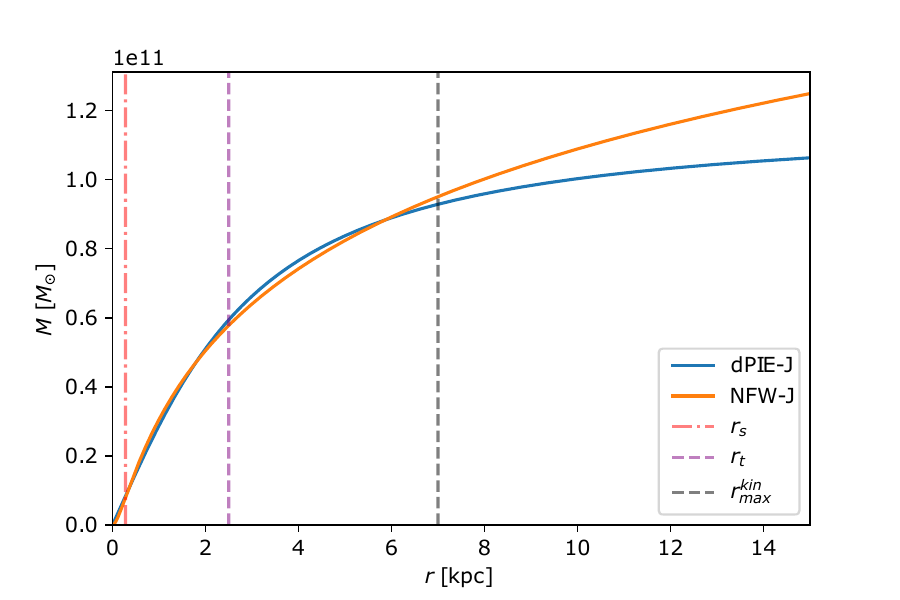}
    \caption{Comparison of the cumulative total mass profiles of a cluster member (ID: 38010) in A2744 obtained with a dPIE-J (cyan) and NFW-J (orange) dynamical models. The inferred median values of $r_t$ and $r_s$ and the maximum radius of the kinematic measurements, $r_{\mathrm{max}}^{\mathrm{kin}}$ are shown as a purple dashed line, a red dash-dotted line and a gray dashed line, respectively.}
    \label{fig:cumulativeM}
\end{figure}

\begingroup
\renewcommand{\arraystretch}{1.07}
\begin{table}
\caption{Comparison of the results of the $\chi^2$ analysis for the three dynamical model parameterizations for the 17 galaxies with more than two measured points in their LOS velocity dispersion profiles.}   
\label{table:chisqu}     
\centering                          
\begin{tabular}{c | c c c}       
\hline\hline                 
Galaxy ID & $\chi^2_{\mathrm{dPIE-J}}$ & $\chi^2_{\mathrm{dPIE-H}}$ & $\chi^2_{\mathrm{NFW-J}}$ \\   
\hline                       
   34423 & $3.10$ & $\boldsymbol{1.33}$ & $8.56$ \\  
   36527 & $0.79$ & $\boldsymbol{0.36}$ & $1.66$ \\
   38010 & $0.49$ & $0.41$ & $\boldsymbol{0.17}$ \\
   38067 & $\boldsymbol{8.74}$ & $9.21$ & $13.31$ \\
   38117 & $3.56$ & $\boldsymbol{3.20}$ & $3.66$ \\
   38729 & $0.03$ & $0.08$ & $\boldsymbol{0.01}$ \\
   39428 & $4.26$ & $\boldsymbol{3.07}$ & $5.07$ \\
   40689 & $\boldsymbol{10.34}$ & $19.21$ & $19.77$ \\
   41950 & $1.18$ & $\boldsymbol{0.91}$ & $1.50$ \\
   42443 & $6.26$ & $\boldsymbol{5.69}$ & $6.85$ \\
   80726 & $36.52$ & $101.08$ & $\boldsymbol{26.07}$ \\
   81838 & $4.42$ & $\boldsymbol{4.15}$ & $4.17$ \\
   81883 & $8.82$ & $9.94$ & $\boldsymbol{0.21}$ \\
   82063 & $\boldsymbol{11.45}$ & $23.85$ & $17.01$ \\
   83064 & $\boldsymbol{2.39}$ & $2.48$ & $3.52$ \\
   85770 & $0.03$ & $0.07$ & $\boldsymbol{0.01}$ \\
   -22 & $\boldsymbol{1.20}$ & $4.69$ & $2.71$ \\
\hline                                  
\end{tabular}
\tablefoot{In the first column, we report the galaxy ID. The subscripts specify the total (dPIE, NFW) and stellar (J: Jaffe, H: Hernquist) mass density distributions used to build each model. The lowest $\chi^2$ value for each galaxy is highlighted.}
\end{table}
\endgroup
The differences in the minimum $\chi^2$ values are typically not large enough to conclude on a clearly preferred profile for the total mass density distribution of the cluster members. Interestingly, the NFW-J profile is typically preferred when the measured LOS velocity dispersion profile is flat or slightly increasing in the first two or three measured points. The NFW density profile inner slope can fit the minor increase in the first and innermost velocity dispersion values, making it better at reproducing the kinematic measurements. Moreover, almost all of the measured profiles for which the NFW parametrization is a better fit are influenced by the rotation of the galaxies. Models that take into account both the rotation and the velocity dispersion could thus strenghten the preference towards the dPIE profile.
Finally, an additional way to compare the dPIE and NFW parameterizations is through the cumulative total mass profile. As one can see in Fig. \ref{fig:cumulativeM}, where we show the cumulative total mass as a function of radial distance from the center of a galaxy, the profiles obtained from the dPIE and NFW parameterizations are substantially identical within the maximum radius where we performed the last kinematic measurement, $r_{\mathrm{max}}^{\mathrm{kin}}$. Similar results apply to all galaxies in our sample. We remark that with deeper MUSE data (i.e., by radially extending the measured kinematic profiles out to larger $r_{\mathrm{max}}^{\mathrm{kin}}$), one could find a preferred total mass density profile for cluster members, distinguishing, for instance, between dPIE and NFW.

\section{The impact of anisotropy on the dynamical models}\label{app:A}
In this Appendix, we discuss the effects of anisotropy in stellar orbits on the measured kinematic profiles and on the posterior probability distributions of the parameters inferred from the dynamical models.

As discussed in App. \ref{app:model}, we modeled the LOS velocity dispersion profiles of cluster members assuming isotropic stellar orbits. This was done by setting $\beta(r)=0\;\forall r$ in Eq. \eqref{sigma_LOS}. This approximation has often been used to describe ETGs, without impacting the reliability of the measurements performed with dynamical models \citep{cappellari2008measuring}. Moreover, the isotropic assumption strongly reduces the computational cost of the dynamical models. On the other hand, assuming a fixed value for $\beta$ is a first order approximation, since ETGs rarely show a constant $\beta$ over their full extension and have been found to prefer orbits that are more tangentially biased $(\beta\leq0)$ towards their centers and more radially biased $(\beta\geq0)$ in the outskirts \citep [see Fig. 10 in][]{cappellari2025early}.
In order to more accurately test the impact of anisotropy on the LOS velocity dispersion profiles and on the $\sigma_0$ values obtained with our dynamical models, we built some alternative dynamical models based on the full Eq. \eqref{sigma_LOS}, with the dPIE-J parametrization and assuming $-0.5\leq\beta(r)\leq0.5,\,\forall r\leq r_{\mathrm{kin}}$, and compared their results with the isotropic dynamical models described in the main body of the paper.
\\

\begin{figure}
    \centering
    \includegraphics[width=\linewidth]{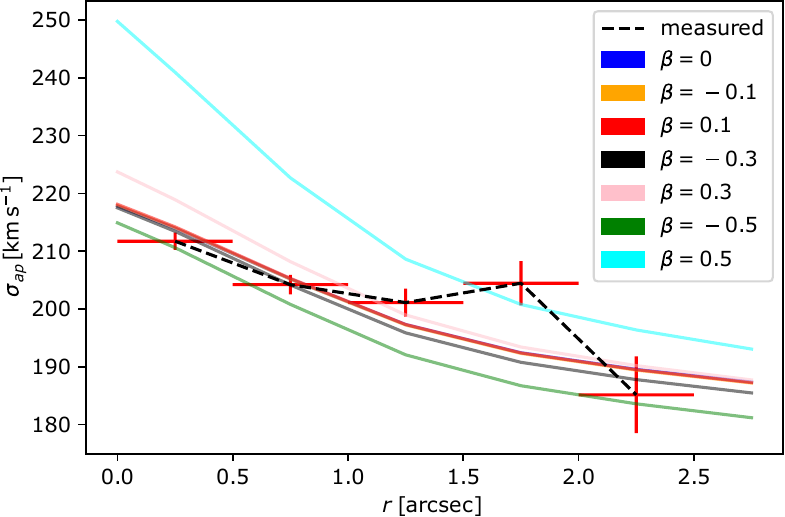}
    \caption{Comparison between the measured and model-predicted LOS velocity dispersion profiles for a cluster member (ID: 80726) in M0416 with fixed $\sigma_0$ and $r_t$ and different anisotropy parameter values. The kinematic measurements are shown as red dots while the predictions of the dynamical models are color-coded according to their $\beta$ value. The scale on the vertical axis is zoomed in, compared to all the other figures in the paper, to show the differences in the profiles that would otherwise be imperceptible.}
    \label{fig:anisotropycomp}
\end{figure}

We compared the measured profiles to those obtained by assuming different constant $\beta(r)=\beta$ values and by fixing $\sigma_0$ and $r_t$ to the median values obtained from the isotropic dPIE-J dynamical models (see Table \ref{table:res17}). In Fig. \ref{fig:anisotropycomp}, we present the comparison between the profiles with $\beta= [0.0, \pm0.1, \pm0.3, \pm0.5]$, color-coded depending on the $\beta$ value, and the kinematic measurements, as red points and error bars, for a cluster member (ID: 80726) in M0416.
From the figure, one can see how the $\beta=0$ and $\beta=\pm0.1$ profiles are almost perfectly overlapping. This implies that assuming a small constant value of anisotropy in the stellar orbits of cluster members does not strongly influence the measured values of $\sigma_0$ and $r_t$. As was shown in Fig. 10 of \citet{cappellari2025early}, this is the typical range of values for $\beta(r)$ at any radial distance from the center of an ETG. 
Moreover, a $\chi^2$ analysis on the predicted profiles with $\beta=0$ and $\beta=\pm0.1$ leads to similar results. From the figure, one can also see how the profiles with more radially biased stellar orbits, namely $\beta=0.3$ and $\beta=0.5$, reproduce the central measured values the least. On the other hand, one can see that the $\beta=-0.3$ profile is similar to the isotropic profile and that assuming extreme tangential anisotropy, such as $\beta=-0.5$, does not strongly influence the central values of the LOS velocity dispersion profiles. These results agree with the fact that radially biased orbits are rarely observed in the central regions of ETGs, while massive ellipticals can show tangentially biased orbits in their centers \citep{binney2011galactic, cappellari2025early}.
\\
\begin{figure}
    \centering
    \includegraphics[width=0.9\linewidth]{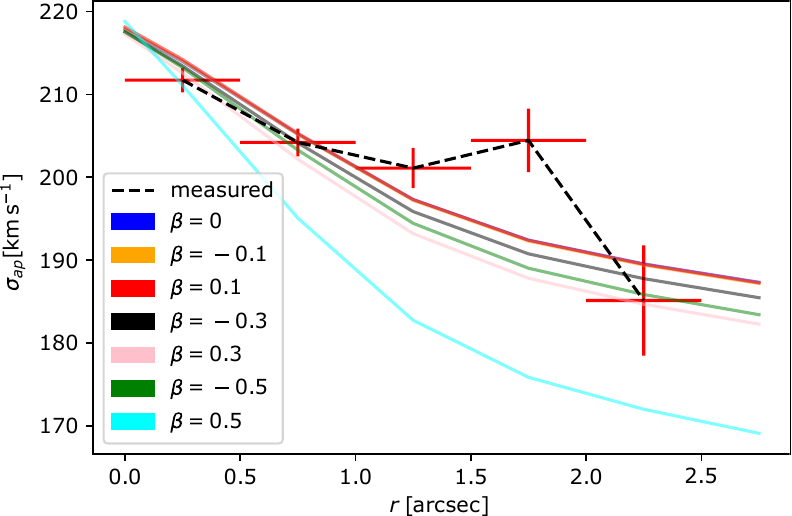}
    \caption{Comparison between the measured and model-predicted LOS velocity dispersion profiles for a cluster member (ID: 80726) in M0416 with fixed $r_t$ and different anisotropy parameter values. The $\sigma_0$ value is manually fixed such that all of the model-predicted profiles show the same central $\sigma$ value. The profiles are color-coded as in Fig. \ref{fig:anisotropycomp}.}
    \label{fig:anisvals}
\end{figure}

In Fig. \ref{fig:anisvals}, we compare the measured profile to those obtained with constant $\beta(r)=\beta$ values (with same color coding as in Fig. \ref{fig:anisotropycomp}), the same fixed $r_t$ as before; and $\sigma_0$ set in order for the profiles obtained with each dynamical model to fit the first kinematic measurement. This is done to probe the effects that different anisotropy values have on the predictions of dynamical models that are forced to reproduce the measured central stellar velocity dispersion. In detail, we set $\sigma_0= 241.9\,\mathrm{km\,s^{-1}}$ (as in Table \ref{table:res17}) for $\beta= [0.0, \pm 0.1, -0.3]$, $\sigma_0=234.9\,\mathrm{km\,s^{-1}}$ for $\beta=0.3$, $\sigma_0=244.9\,\mathrm{km\,s^{-1}}$ for $\beta=-0.5$ and $\sigma_0=211.9\,\mathrm{km\,s^{-1}}$ for $\beta=0.5$. From the figure, it is evident that the profiles with $\beta=0, \pm 0.1$ accurately reproduce the kinematic measurements, while the dynamical models worsen at reproducing the measured profiles for increasing $|\beta|$ values. To avoid this, one should assign increasingly higher values of $r_t$ at increasingly larger values of $|\beta|$. A $\chi^2$ analysis on the LOS velocity dispersion profiles shows that the $\beta=0$ and $\beta=\pm 0.1$ parameterizations, as before, are the ones that better reproduce the kinematic measurements. Moreover, comparing the $\sigma_0$ values we manually set for each dynamical model, one can see how assuming low and plausible values of anisotropy does not require changing the values of the kinematic parameters in order for the model to reproduce the measured profile. This implies that the dynamical models are able to infer the values of $\sigma_0$ and $r_t$, independently of the anisotropy, as long as the assumptions on $\beta$ are realistic.
\\
\begin{figure}
    \centering
    \includegraphics[width=0.9\linewidth]{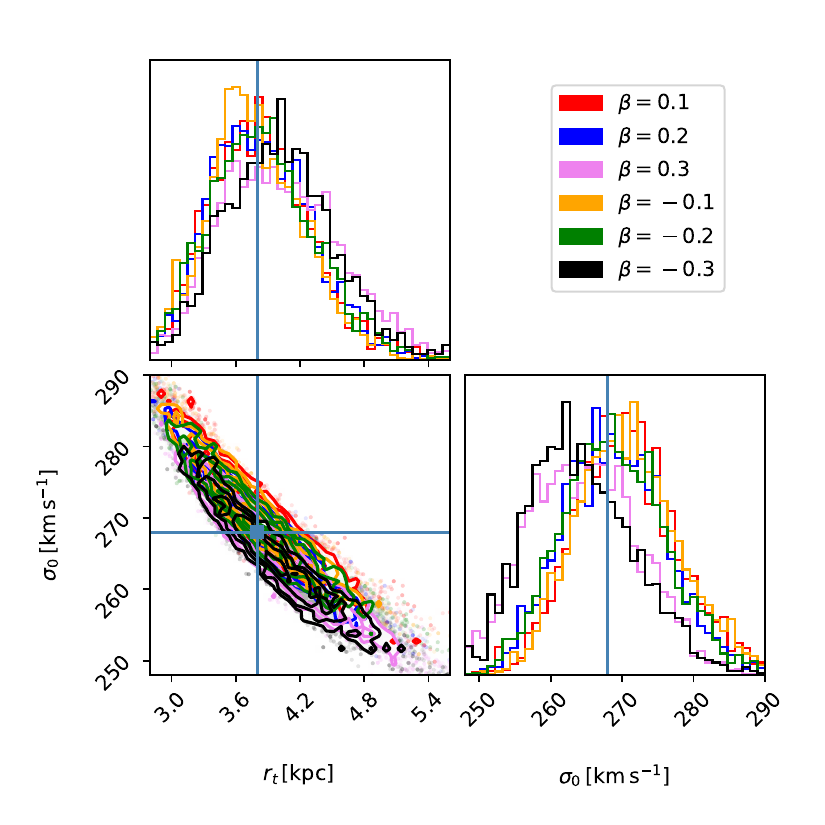}
    \caption{Comparison of the $\sigma_0$ and $r_t$ posterior probability distributions obtained with different anisotropy values.
    We plot the values of the two parameters obtained from the isotropic dPIE-J dynamical model as blue lines. Each posterior distribution is color-coded, depending on the value of the anisotropy parameter assumed in the inference procedure.}
    \label{fig:betapost}
\end{figure}
Finally, we developed a full dPIE-J anisotropic dynamical model to test the effects of anisotropy on the inferred values of $\sigma_0$ and $r_t$. In this model, $\beta$ is constant at all radii and the MCMC optimization process (see Sect. \ref{ss:MCMC}) is performed on $\sigma_0$ and $r_t$ with priors defined as in the main body of this paper. Given that a full model has a higher computational cost than an isotropic model, we adopted MCMC chains of 2000 steps each, with initial $\sigma_0$ and $r_t$ values set as the values obtained from the isotropic models (see Table \ref{table:res17}). We ran the dynamical model assuming six different values of the anisotropy parameter, namely $\beta=[\pm0.1,\pm0.2,\pm0.3]$, and compared the resulting posterior probability distributions with those of the isotropic dynamical models.
The comparison of the posterior probability distributions of $\sigma_0$ and $r_t$ for one galaxy (ID: 83064) in M0416 is shown in Fig. \ref{fig:betapost}. The median values obtained with the isotropic model are shown as blue lines. From the figure, it is evident that all the posterior distributions are similar and that the inferred median values are consistent with the reference isotropic model, regardless of the assumed $\beta$ value. Our analysis demonstrates that assuming isotropy or small and constant anisotropy parameter values to describe ETG stellar orbits does not significantly impact the predictions of our dynamical models. This result confirms the robustness of our modeling methods and measurements, thereby strengthening the conclusions of this paper. Similarly to us, \citet{Liang2025} find that assuming isotropic profiles or the Osipkov-Merritt \citep{osipkov1979spherical, merritt1985spherical} model to define $\beta(r)$ does not impact the inferred galaxy mass density distributions.
In future works, more complex dynamical models could consider a full functional form for the anisotropy parameter $\beta(r)$ or define $\beta$ as a free parameter to be inferred.

\section{Sample and main physical parameters}\label{app:B}
In this Appendix, we present our sample of cluster members, including their redshift and structural parameters, as described in \citet{tortorelli2023kormendy} for the galaxies in A2744 and M0416, respectively. A further analysis was presented in \citet{granata2026velocity}. We also present the relevant measurements that were used to build the dynamical models and calibrate the scaling relations discussed in the main body of the paper.
\begin{table*}
\caption{Relevant measurements and values for the 17 cluster members with more than two measurements in their LOS velocity dispersion profiles.}     
\label{table:res17}     
\centering                          
\begin{tabular}{c c c c| c c c |c c c | c }       
\hline
\hline
\multicolumn{11}{c}{A2744}\\
\hline 
\multicolumn{4}{c|}{} & \multicolumn{3}{c|}{dPIE-J} & \multicolumn{3}{c|}{NFW-J}& \multicolumn{1}{c}{} \\   
\hline 
    Gal ID & $z$ & $m_{\mathrm{F160W}}$ & $R_e^{\mathrm{F814W}}$& $\sigma_0$& $r_t$& $v_{\text{max, dPIE}}^{\text{circ}}\,$ & $\rho_0$ & $r_s$ & $v_{\text{max, NFW}}^{\text{circ}}$ & $z_{\mathrm{score}}$\\
   & & &$[\mathrm{kpc}]$ & $[\mathrm{km\,s^{-1}}]$ & $[\mathrm{kpc}]$ & $[\mathrm{km\,s^{-1}}]$ & $[\mathrm{10^9\, M_{\odot}\,kpc^{-3}}]$ & $[\mathrm{kpc}]$ & $[\mathrm{km\,s^{-1}}]$ & \\
\hline
   $\boldsymbol{34423}$ & $0.303$ & $17.68$ & $3.53$ &$216.3^{+3.4}_{-2.1}$ & $37.5^{+79.7}_{-17.0}$ & $305.9^{+3.9}_{-3.9}$ & $1.7^{+0.2}_{-0.2}$ & $3.0^{+0.7}_{-0.7}$ & $318.6^{+2.4}_{-2.4}$ & $2.71$\\  
   $\boldsymbol{36527}$ & $0.316$ & $18.08$ & $2.75$ &$272.4^{+6.0}_{-6.6}$ & $13.1^{+9.3}_{-3.2}$ & $385.4^{+8.6}_{-8.6}$ & $1.2^{+0.2}_{-0.2}$ & $8.6^{+3.2}_{-2.7}$ & $394.6^{+5.4}_{-5.4}$ & $0.91$ \\
   $\boldsymbol{38010}$ & $0.317$ & $18.76$ & $1.53$ & $255.0^{+16.6}_{-14.1}$ & $2.5^{+0.6}_{-0.5}$ & $361.7^{+22.5}_{-22.5}$ & $150^{+280}_{-100}$ & $0.3^{+0.2}_{-0.1}$ & $364.8^{+40.9}_{-40.9}$ & $0.07$\\
   $\boldsymbol{38067}$ & $0.317$ & $18.18$ & $3.24$ & $223.0^{+2.1}_{-1.9}$ & Flat & $315.4^{+2.9}_{-2.9}$ & $1.6^{+0.8}_{-0.5}$ & $2.4^{+0.4}_{-0.4}$ & $328.9^{+3.1}_{-3.1}$ & $3.20$\\
   $\boldsymbol{38117}$ & $0.299$ & $18.60$ & $1.71$ & $217.4^{+9.3}_{-10.2}$ & $8.4^{+14.9}_{-2.6}$ & $307.1^{+14.4}_{-14.4}$ & $22.0^{+28.0}_{-9.6}$ & $0.7^{+0.2}_{-0.2}$ & $326.7^{+16.8}_{-16.8}$ & $0.89$ \\
   $\boldsymbol{38729}$ & $0.319$ & $19.20$ & $2.83$& $149.3^{+2.7}_{-2.6}$ & Flat & $211.1^{+3.7}_{-3.7}$ & $4.2^{+9.2}_{-2.8}$ & $1.0^{+0.7}_{-0.4}$ & $226.6^{+11.6}_{-11.6}$ & $1.27$ \\
   $\boldsymbol{39428}$ & $0.300$ & $18.20$ & $3.13$ & $192.6^{+4.9}_{-5.3}$ & $9.8^{+5.5}_{-2.1}$ & $272.4^{+6.9}_{-6.9}$ & $7.1^{+3.3}_{-2.2}$ & $1.0^{+0.2}_{-0.2}$ & $279.5^{+5.2}_{-5.2}$ & $0.82$\\
   $\boldsymbol{40689}$ & $0.301$ & $17.02$ & $7.26$ & $252.0^{+2.0}_{-2.0}$ & $23.1^{+4.4}_{-3.2}$ & $357.2^{+2.8}_{-2.8}$ & $2.4^{+0.3}_{-0.3}$ & $2.2^{+0.1}_{-0.1}$ & $365.7^{+1.8}_{-1.8}$ & $2.55$\\
   $\boldsymbol{41950}$ & $0.317$ & $18.44$ & $1.94$& $244.7^{+7.7}_{-8.2}$ & $11.1^{+22.0}_{-3.4}$ & $346.3^{+11.0}_{-11.0}$ & $14.0^{+8.4}_{-5.2}$ & $0.9^{+0.2}_{-0.2}$ & $359.9^{+9.2}_{-9.2}$ & $0.94$\\
   $\boldsymbol{42443}$ & $0.303$ & $18.77$ & $2.56$& $114.1^{+2.4}_{-2.0}$ & Flat & $161.3^{+3.1}_{-3.1}$ & $6.3^{+11.0}_{-3.3}$ & $0.7^{+0.3}_{-0.2}$ & $182.3^{+11.8}_{-11.8}$ & $1.73$ \\
   \hline 
    \multicolumn{11}{c}{M0416}\\
    \hline
\multicolumn{4}{c|}{} & \multicolumn{3}{c|}{dPIE-J} & \multicolumn{3}{c|}{NFW-J}& \multicolumn{1}{c}{} \\   
\hline 
    Gal ID & $z$ & $m_{\mathrm{F160W}}$ & $R_e^{\mathrm{F814W}}$& $\sigma_0$& $r_t$& $v_{\text{max, dPIE}}^{\text{circ}}\,$ & $\rho_0$ & $r_s$ & $v_{\text{max, NFW}}^{\text{circ}}$ & $z_{\mathrm{score}}$\\
   & & &$[\mathrm{kpc}]$ & $[\mathrm{km\,s^{-1}}]$ & $[\mathrm{kpc}]$ & $[\mathrm{km\,s^{-1}}]$ & $[\mathrm{10^9\, M_{\odot}\,kpc^{-3}}]$ & $[\mathrm{kpc}]$ & $[\mathrm{km\,s^{-1}}]$ & \\
\hline
   $\boldsymbol{80726}$ & $0.402$ & $17.78$ & $5.09$& $241.9^{+1.1}_{-1.1}$ & Flat & $342.0^{+1.6}_{-1.6}$ & $1.2^{+0.2}_{-0.1}$ & $3.2^{+0.2}_{-0.2}$ & $373.2^{+1.8}_{-1.8}$ & $12.82$\\
   81838 & $0.396$ & $19.08$ & $1.11$& $248.3^{+11.6}_{-8.0}$ & $12.4^{+88.9}_{-5.5}$ & $352.0^{+13.8}_{-13.8}$ & $44.0^{+91.0}_{-18.0}$ & $0.6^{+0.1}_{-0.2}$ & $393.2^{+22.6}_{-22.6}$ & $1.55$ \\
   $\boldsymbol{81883}$ & $0.406$ & $19.44$ & $1.85$& $177.9^{+1.9}_{-1.8}$ & Flat & $251.7^{+2.6}_{-2.6}$ & $0.5^{+0.2}_{-0.2}$ & $3.4^{+0.7}_{-0.6}$ & $269.9^{+5.7}_{-5.7}$ & $2.93$ \\
   $\boldsymbol{82063}$ & $0.395$ & $17.96$ & $4.37$& $221.3^{+2.6}_{-2.6}$ & $18.6^{+5.5}_{-3.3}$ & $312.9^{+3.7}_{-3.7}$ & $3.7^{+0.9}_{-0.7}$ & $1.6^{+0.2}_{-0.2}$ & $330.1^{+2.5}_{-2.5}$ & $3.87$ \\
   $\boldsymbol{83064}$ & $0.397$ & $18.91$ & $2.83$& $267.7^{+7.9}_{-6.9}$ & $3.9^{+0.5}_{-0.5}$ & $378.7^{+10.5}_{-10.5}$ & $110^{+70}_{-32}$ & $0.4^{+0.1}_{-0.1}$ & $404.3^{+16.1}_{-16.1}$ & $1.33$ \\
   $\boldsymbol{85770}$ & $0.406$ & $19.59$ & $1.78$& $164.9^{+6.1}_{-4.9}$ & $13.5^{+58.3}_{-5.2}$ & $233.0^{+7.8}_{-7.8}$ & $5.8^{+4.2}_{-2.4}$ & $0.9^{+0.3}_{-0.2}$ & $245.4^{+6.4}_{-6.4}$ & $1.23$\\
   $\boldsymbol{-22}$ & $0.397$ & $17.79$ & $4.64$& $261.0^{+3.5}_{-2.6}$ & Flat & $368.8^{+3.9}_{-3.9}$ & $1.9^{+0.6}_{-0.5}$ & $2.6^{+0.4}_{-0.3}$ & $380.4^{+3.5}_{-3.4}$ & $2.21$\\
\hline                                  
\end{tabular}
\tablefoot{We report each galaxy ID, redshift, apparent magnitude in the F160W band, and circularized effective radius in the F814W band that were presented in \citet{tortorelli2023kormendy} and \citet{granata2026velocity}. We also list the values of the parameters $\sigma_0$ and $r_t$ inferred with the dPIE-J dynamical models; and $\rho_0$ and $r_s$ inferred with the NFW-J dynamical models. The maximum circular velocities are indicated with $v_{\text{max}}^{\text{circ}}$, with the $z_{\mathrm{score}}$ to compare their distributions. We highlight the ID of the galaxies in the `RS' sample.}
\end{table*}

\begin{table*}
\caption{Relevant measurements and values for the galaxies with less than three measured points in their LOS velocity dispersion profiles.}     
\label{table:res2p}     
\centering                          
\begin{tabular}{c c c c c| c c c c c }   
\hline
\hline 
\multicolumn{10}{c}{A2744}\\
\hline
Gal ID & $z$ & $m_{\mathrm{F160W}}$ & $R_e^{\mathrm{F814W}}\,[\mathrm{kpc}]$ & $\sigma_0\,[\mathrm{km\,s^{-1}}]$ & Gal ID & $z$ & $m_{\mathrm{F160W}}$ & $R_e^{\mathrm{F814W}}\,[\mathrm{kpc}]$ & $\sigma_0\,[\mathrm{km\,s^{-1}}]$\\
\hline 
$\boldsymbol{32547}$ & $0.320$ & $19.33$ & $2.39$& $113.9^{+2.4}_{-2.4}$ & $\boldsymbol{33540}$ & $0.322$ & $19.56$ & $1.86$& $139.3^{+3.4}_{-3.0}$\\
    $\boldsymbol{34556}$ & $0.319$ & $19.57$ & $0.75$& $242.8^{+4.8}_{-4.5}$ & $\boldsymbol{35339}$ & $0.316$ & $19.21$ & $1.98$& $171.8^{+2.6}_{-2.7}$\\
    $\boldsymbol{35693}$ & $0.299$ & $19.97$ & $0.57$& $203.0^{+7.0}_{-4.4}$ & $\boldsymbol{36210}$ & $0.315$ & $17.87$ & $3.04$& $271.0^{+1.9}_{-1.9}$\\
    $\boldsymbol{37068}$ & $0.320$ & $19.41$ & $0.45$& $308.9^{+6.7}_{-6.7}$ & $\boldsymbol{37229}$ & $0.304$ & $19.79$ & $0.61$& $226.8^{+4.0}_{-3.7}$\\
    $\boldsymbol{37230}$ & $0.320$ & $20.17$ & $1.43$& $115.0^{+3.2}_{-3.0}$ & $\boldsymbol{37344}$ & $0.319$ & $18.68$ & $2.50$& $195.3^{+2.4}_{-2.4}$\\
    37824 & $0.300$ & $17.24$ & $3.79$& $367.6^{+2.0}_{-2.1}$ & $\boldsymbol{37954}$ & $0.323$ & $18.56$ & $2.64$& $202.2^{+2.6}_{-2.1}$\\
    $\boldsymbol{38930}$ & $0.302$ & $18.74$ & $1.21$& $192.2^{+1.7}_{-1.5}$ & $\boldsymbol{39072}$ & $0.316$ & $18.47$ & $7.39$& $118.4^{+2.3}_{-2.1}$\\
    $\boldsymbol{39503}$ & $0.300$ & $19.49$ & $1.35$& $159.0^{+3.0}_{-2.8}$ & $\boldsymbol{39646}$ & $0.319$ & $19.47$ & $3.27$& $160.3^{+5.3}_{-5.3}$\\
    $\boldsymbol{39876}$ & $0.294$ & $20.08$ & $1.03$& $112.4^{+4.2}_{-3.8}$ & $\boldsymbol{40270}$ & $0.316$ & $19.79$ & $0.84$& $205.5^{+5.2}_{-4.7}$\\
    $\boldsymbol{40314}$ & $0.297$ & $19.10$ & $1.51$& $186.5^{+2.3}_{-2.2}$ & $\boldsymbol{40478}$ & $0.297$ & $18.78$ & $4.58$& $111.8^{+14.1}_{-3.0}$\\
    $\boldsymbol{40884}$ & $0.302$ & $19.68$ & $1.44$ & $153.6^{+2.9}_{-2.9}$ & $\boldsymbol{41303}$ & $0.303$ & $19.42$ & $3.47$& $84.3^{+3.4}_{-3.4}$\\
    $\boldsymbol{41418}$ & $0.316$ & $19.51$ & $0.80$& $256.0^{+4.6}_{-4.5}$ & $\boldsymbol{42149}$ & $0.303$ & $19.60$ & $1.15$& $161.6^{+3.0}_{-2.9}$\\
    $\boldsymbol{42269}$ & $0.303$ & $19.86$ & $1.20$& $139.6^{+3.3}_{-3.1}$ & $\boldsymbol{21367^*}$ & $0.321$ & $19.85$ & $2.04$& $102.4^{+3.5}_{-3.5}$\\
    $\boldsymbol{32768^*}$ & $0.302$ & $20.64$ & $1.51$& $83.5^{+6.9}_{-6.9}$ & $\boldsymbol{33328^*}$ & $0.299$ & $19.83$ & $3.28$& $102.1^{+5.5}_{-4.7}$\\
    $\boldsymbol{33410^*}$ & $0.315$ & $21.16$ & $1.72$& $102.8^{+10.5}_{-10.9}$ & $\boldsymbol{33699^*}$ & $0.319$ & $20.04$ & $1.10$& $157.9^{+6.1}_{-5.8}$\\
    $\boldsymbol{33870^*}$ & $0.320$ & $20.53$ & $0.42$& $192.0^{+7.3}_{-7.0}$ & $\boldsymbol{34439^*}$ & $0.318$ & $20.85$ & $0.39$& $138.3^{+6.5}_{-5.9}$\\
    $\boldsymbol{35576^*}$ & $0.300$ & $21.22$ & $0.56$& $103.1^{+6.7}_{-6.3}$ & $\boldsymbol{35908^*}$ & $0.304$ & $20.48$ & $0.94$& $141.6^{+4.9}_{-4.7}$\\
    $\boldsymbol{36043^*}$ & $0.316$ & $20.66$ & $0.93$& $118.9^{+6.0}_{-4.9}$ & $\boldsymbol{36220^*}$ & $0.316$ & $20.25$ & $1.74$& $98.0^{+4.7}_{-4.5}$\\
    $36298^*$ & $0.316$ & $21.19$ & $1.23$& $74.1^{+6.0}_{-6.4}$ & $\boldsymbol{36849^*}$ & $0.317$ & $21.21$ & $0.91$& $90.5^{+6.8}_{-6.6}$\\
    $\boldsymbol{36953^*}$ & $0.314$ & $20.95$ & $0.59$& $130.2^{+9.9}_{-8.3}$ & $\boldsymbol{37542^*}$ & $0.324$ & $21.89$ & $1.14$& $77.3^{+12.8}_{-11.1}$\\
    $\boldsymbol{37609^*}$ & $0.305$ & $20.60$ & $1.27$& $88.8^{+7.2}_{-7.1}$ & $\boldsymbol{37825^*}$ & $0.299$ & $20.89$ & $1.56$& $94.5^{+8.8}_{-8.3}$\\
    $\boldsymbol{38143^*}$ & $0.303$ & $20.79$ & $1.38$& $108.1^{+7.2}_{-6.9}$ & $38252^*$ & $0.298$ & $21.46$ & $0.97$& $83.2^{+7.7}_{-7.2}$\\
    $\boldsymbol{38275^*}$ & $0.312$ & $20.08$ & $3.00$& $78.6^{+4.2}_{-4.0}$ & $\boldsymbol{39283^*}$ & $0.306$ & $20.17$ & $1.65$& $107.2^{+5.2}_{-5.0}$\\
    $\boldsymbol{40032^*}$ & $0.297$ & $20.75$ & $1.46$& $78.4^{+5.2}_{-5.5}$ & $40428^*$ & $0.295$ & $20.81$ & $1.73$& $91.4^{+8.8}_{-7.9}$\\
    $\boldsymbol{40832^*}$ & $0.312$ & $20.03$ & $2.09$& $100.3^{+6.0}_{-5.5}$ & $\boldsymbol{44545^*}$ & $0.318$ & $20.91$ & $0.76$& $121.5^{+8.4}_{-8.0}$\\
\hline 
\multicolumn{10}{c}{M0416}\\
\hline
    Gal ID & $z$ & $m_{\mathrm{F160W}}$ & $R_e^{\mathrm{F814W}}\,[\mathrm{kpc}]$ & $\sigma_0\,[\mathrm{km\,s^{-1}}]$ & Gal ID & $z$ & $m_{\mathrm{F160W}}$ & $R_e^{\mathrm{F814W}}\,[\mathrm{kpc}]$ & $\sigma_0\,[\mathrm{km\,s^{-1}}]$\\
\hline
    $\boldsymbol{79058}$ & $0.400$ & $18.02$ & $5.44$& $227.5^{+3.8}_{-3.4}$ & $\boldsymbol{81285}$ & $0.401$ & $18.58$ & $2.26$& $322.7^{+33.3}_{-20.1}$\\
    $\boldsymbol{82024}$ & $0.399$ & $20.64$ & $1.08$& $119.9^{+2.5}_{-2.5}$ & $\boldsymbol{83948}$ & $0.394$ & $20.27$ & $1.75$& $131.6^{+2.5}_{-2.4}$ \\
    $\boldsymbol{84025}$ & $0.398$ & $19.76$ & $0.74$& $238.8^{+2.7}_{-2.6}$ & $\boldsymbol{84536}$ & $0.406$ & $19.90$ & $1.33$& $173.2^{+2.5}_{-2.5}$\\
    $\boldsymbol{85357}$ & $0.397$ & $20.77$ & $0.64$& $131.1^{+3.3}_{-3.6}$ & $\boldsymbol{86038}$ & $0.395$ & $20.43$ & $1.41$& $122.1^{+2.4}_{-2.2}$\\
    $\boldsymbol{77317}$& $0.392$ & $20.03$ & $0.83$& $196.7^{+6.7}_{-5.4}$ & $\boldsymbol{79978}$ & $0.402$ & $19.77$ & $1.39$& $170.2^{+3.8}_{-3.4}$\\
    $\boldsymbol{80310}$ & $0.404$ & $19.55$ & $1.22$& $168.2^{+4.4}_{-3.4}$ & $\boldsymbol{80466}$ & $0.405$ & $19.92$ & $0.71$& $263.9^{+6.0}_{-5.5}$\\
    $\boldsymbol{82442}$ & $0.392$ & $19.49$ & $1.19$& $228.8^{+3.8}_{-3.6}$ & $\boldsymbol{-8^*}$ & $0.398$ & $21.84$ & $0.67$& $101.5^{+8.5}_{-7.0}$\\
    $\boldsymbol{-15^*}$ & $0.391$ & $20.84$ & $0.91$& $101.3^{+7.2}_{-6.5}$ & $-104^*$ & $0.400$ & $22.30$ & $1.55$& $140.9^{+9.9}_{-9.4}$\\
    $\boldsymbol{-144^*}$ & $0.401$ & $20.90$ & $2.01$& $85.8^{+7.1}_{-7.0}$ & $-157^*$ & $0.398$ & $21.61$ & $1.59$ & $147.9^{+20.6}_{-19.8}$\\
    $\boldsymbol{78004^*}$ & $0.392$ & $21.71$ & $0.69$& $131.4^{+14.7}_{-13.8}$ & $\boldsymbol{78230^*}$ & $0.400$ & $21.03$ & $0.99$& $94.2^{+8.3}_{-7.2}$\\
    $\boldsymbol{79635^*}$ & $0.397$ & $21.35$ & $0.48$& $153.0^{+11.7}_{-11.0}$ & $79898^*$ & $0.391$ & $21.06$ & $2.41$& $87.5^{+6.3}_{-6.3}$\\
    $79957^*$ & $0.401$ & $21.49$ & $1.41$& $86.9^{+9.6}_{-9.3}$ & $\boldsymbol{80895^*}$ & $0.395$ & $20.89$ & $1.08$& $120.6^{+8.6}_{-8.2}$\\
    $\boldsymbol{81125^*}$ & $0.397$ & $21.49$ & $0.42$& $130.2^{+9.4}_{-9.3}$ & $\boldsymbol{81561^*}$ & $0.400$ & $21.01$ & $0.70$& $133.0^{+7.4}_{-6.7}$\\
    $\boldsymbol{81782^*}$ & $0.393$ & $20.54$ & $1.78$& $110.0^{+10.7}_{-9.2}$ & $82012^*$ & $0.404$ & $22.29$ & $0.84$& $72.4^{+10.5}_{-10.0}$\\
    $\boldsymbol{82042^*}$ & $0.389$ & $19.94$ & $1.02$& $150.6^{+5.6}_{-4.9}$ & $82585^*$ & $0.401$ & $22.99$ & $1.64$& $61.9^{+17.4}_{-16.8}$\\
    $\boldsymbol{82765^*}$ & $0.391$ & $19.50$ & $1.91$& $118.1^{+2.2}_{-1.8}$ & $\boldsymbol{82863^*}$ & $0.400$ & $20.13$ & $0.51$& $283.9^{+9.2}_{-8.3}$\\
    $\boldsymbol{82961^*}$ & $0.395$ & $21.42$ & $0.86$& $119.9^{+6.0}_{-5.9}$ & $\boldsymbol{83261^*}$ & $0.402$ & $20.34$ & $1.84$& $135.6^{+6.4}_{-6.3}$\\
    $83541^*$ & $0.402$ & $21.21$ & $2.63$& $70.1^{+5.9}_{-5.8}$ & $\boldsymbol{83625^*}$ & $0.392$ & $21.86$ & $0.85$& $100.5^{+7.5}_{-5.8}$\\
    $\boldsymbol{84326^*}$ & $0.400$ & $20.88$ & $0.41$& $174.9^{+11.6}_{-11.7}$ & $85042^*$ & $0.394$ & $22.09$ & $1.49$& $42.9^{+16.0}_{-14.6}$\\
    $85339^*$ & $0.389$ & $22.49$ & $0.86$& $66.4^{+11.5}_{-11.4}$ & $\boldsymbol{85556^*}$ & $0.395$ & $21.15$ & $0.67$& $117.1^{+4.5}_{-4.3}$\\
    $\boldsymbol{85812^*}$ & $0.398$ & $21.09$ & $1.91$& $87.6^{+3.3}_{-3.4}$ & $\boldsymbol{85939^*}$ & $0.389$ & $21.26$ & $1.46$& $79.8^{+11.0}_{-9.4}$\\
\hline
\end{tabular}
\tablefoot{We report each galaxy ID, redshift, apparent magnitude in the F160W band, circularized effective radius in the F814W band that were presented in \citet{tortorelli2023kormendy} and \citet{granata2026velocity}, and central stellar velocity dispersion obtained with the dPIE-J dynamical models. We do not report the $r_t$ values since the kinematic profiles are not radially extended enough to accurately measure them. We highlight the IDs of the galaxies in the `RS' sample and mark with a star ($^*$) those with only one measurement in the LOS velocity dispersion profile.}
\end{table*}

\end{appendix}
\end{document}